\documentclass[superscriptaddress,twocolumn,
 amsmath,
 amssymb,
 prl,
]{revtex4-2}
\usepackage{soul}
\usepackage{slashed}
\usepackage{xcolor}
\usepackage{graphicx}
\usepackage{dcolumn}
\usepackage{bm}
\usepackage{hyperref}
\usepackage[mathlines]{lineno}

\begin{document}


\title{Information field based global Bayesian inference of the jet transport coefficient}

\author{Man Xie}
\affiliation{Key Laboratory of Quark and Lepton Physics (MOE) \& Institute of Particle Physics, Central China Normal University, Wuhan 430079, China}

\author{Weiyao Ke}
\email[]{weiyaoke@lanl.gov}
\affiliation{Department of Physics, University of California, Berkeley, California 94720, USA}
\affiliation{Nuclear Science Division MS 70R0319, Lawrence Berkeley National Laboratory, Berkeley, California 94720, USA}
\affiliation{Theoretical Division, Los Alamos National Laboratory, Los Alamos, NM 87545, USA}

\author{Hanzhong Zhang}
\email[]{zhanghz@mail.ccnu.edu.cn}
\affiliation{Key Laboratory of Quark and Lepton Physics (MOE) \& Institute of Particle Physics, Central China Normal University, Wuhan 430079, China}

\author{Xin-Nian Wang}
\email[]{xnwang@lbl.gov}
\affiliation{Department of Physics, University of California, Berkeley, California 94720, USA}
\affiliation{Nuclear Science Division MS 70R0319, Lawrence Berkeley National Laboratory, Berkeley, California 94720, USA}

\date{\today}
\begin{abstract}
Bayesian statistical inference is a powerful tool for model-data comparisons and extractions of physical parameters that are often unknown functions of system variables. Existing Bayesian analyses often rely on explicit parametrizations of the unknown function. It can introduce long-range correlations that impose fictitious constraints on physical parameters in regions of the variable space that are not probed by the experimental data. We develop an information field (IF) approach to modeling the prior distribution of the unknown function that is free of long-range correlations. We apply the IF approach to the first global Bayesian inference of the jet transport coefficient $\hat q$ as a function of temperature ($T$) from all existing experimental data on single-inclusive hadron, di-hadron and $\gamma$-hadron spectra in heavy-ion collisions at RHIC and LHC energies. The extracted $\hat q/T^3$ exhibits a strong $T$-dependence as a result of the progressive constraining power when data from more central collisions and at higher colliding energies are incrementally included. 
The IF method guarantees that the extracted $T$-dependence is not biased by a specific functional form.
\end{abstract}

\maketitle


\noindent {\color{blue}\it Introduction--} Recent developments in Bayesian statistical inference have brought fruitful advances in physics research and greatly improved our ability to constrain complex physical models with ever increasing data. Some prime examples in high-energy nuclear physics are the nominal study of the equation of state of the quark-gluon plasma (QGP) \cite{Pratt:2015zsa}, simultaneous studies of initial conditions of nuclear collisions and QGP shear and bulk viscosity \cite{Bernhard:2015hxa,PhysRevC.94.024907,Bernhard:2019bmu,JETSCAPE:2020mzn,JETSCAPE:2020shq,Nijs:2020roc,Nijs:2020ors,Parkkila:2021tqq,Parkkila:2021yha}, heavy quark diffusion parameter \cite{Xu:2017obm,PhysRevC.98.064901,Liu:2021dpm} and jet transport coefficient for parton energy loss in the QGP \cite{Ke:2020clc,JETSCAPE:2021ehl}. 
The pivot of these analyses is the Bayes' Theorem: ``Posterior $=$ Prior $\times$ Likelihood''. It describes how a prior belief of the distribution of certain physical parameters will be updated to the posterior distribution after comparing theory with new experimental data, as encoded in the likelihood function that quantifies the model's descriptive power. 

Many physical quantities to be inferred from data are unknown functions of system variables. For example, viscosity depends on temperature and chemical potential of the QGP. The jet transport coefficient depends on temperature and parton energy \cite{Casalderrey-Solana:2007xns,He:2015pra,JETSCAPE:2021ehl}. The nuclear parton distributions can be functions of the parton momentum fraction, the renormalization scale and the impact-parameter~\cite{PhysRevLett.120.152502,DelDebbio:2021whr}. In all recent analyses, these unknown functions are parameterized in intuitive forms either for convenience or motivated by physical insights. Bayesian inferences are then performed to extract these parameters from model-data comparisons. 
A major drawback of explicit parameterizations is that they often introduce unwarranted long-range correlations in the prior distributions of physical functions between different regions of the variable space, where different experimental data sets are supposed to provide independent constraints.
For example, data in peripheral heavy-ion collisions and at lower colliding energies, which only probe QGP at low temperatures, can lead to fictitious constraints on physical quantities at higher temperatures as predetermined by the parameterizations.

In this work, we design a framework for Bayesian inference of unknown functions based on the idea of the information field (IF)  \cite{PhysRevLett.77.4693,Ensslin:2013ji,https://doi.org/10.48550/arxiv.physics/9912005}. The advantage of the IF approach is that it provides a non-parametric representation of the unknown functional space and eliminates prior bias and unnecessary long-range correlations in the prior. 
It can be easily generalized to higher-dimensional functional inference. Furthermore, it facilitates sensitivity analyses that map the constraining power of the experimental data to the variable space of physical functions. As a proof of principle, we apply this IF approach to the Bayesian inference of the jet transport coefficient $\hat{q}(T)$ from experimental data on jet quenching, with the temperature $T$ being the system variable.

Jet quenching is the suppression of large transverse momentum ($p_{\rm T}$) jets and hadrons caused by energy loss of energetic partons (quarks and gluons) as they propagate through QGP in high-energy heavy-ion (A+A) collisions \cite{Gyulassy:1990ye,Wang:1991xy,Wang:1998bha,Majumder:2010qh,Qin:2015srf}. Parton energy loss is controlled by the jet transport coefficient $\hat q$, defined as the transverse momentum broadening squared per unit length 
\cite{Baier:1996kr,Baier:1996sk,Zakharov:1996fv,Wiedemann:2000za,Guo:2000nz, Wang:2001ifa,Majumder:2009ge}. In QGP, it can depend on both the local temperature along the jet propagation path and jet energy. Past efforts in extracting $\hat q$ from model comparisons to experimental data relied on simple fits \cite{JET:2013cls,Liao:2008dk,Xu:2014tda} or Bayesian inferences from single inclusive hadron or jet spectra \cite{JETSCAPE:2021ehl,Ke:2020clc}.
We will carry out the first global Bayesian inference of $\hat q$ with the IF approach from combined data on suppression of single inclusive hadron \cite{Adare:2008qa,Adare:2012wg,CMS:2012aa,Abelev:2012hxa,Aad:2015wga,Khachatryan:2016odn,Acharya:2018qsh}, $\gamma$-hadron \cite{Abelev:2009gu,STAR:2016jdz}, and di-hadron \cite{Abelev:2009gu,STAR:2016jdz,Aamodt:2011vg,Adam:2016xbp,Conway:2013xaa} spectra at both the Relativistic Heavy-ion Collider (RHIC) and the Large Hadron Collider (LHC) energies.  We will exclude data on reconstructed jets in this study, since they are also sensitive to the modeling of jet-induced medium response~\cite{He:2015pra,PhysRevC.95.044909,Chen:2017zte,KunnawalkamElayavalli:2017hxo,Milhano:2017nzm,He:2018xjv}. 

\noindent {\color{blue}\em NLO parton model with energy loss--}
We use the next-leading-order (NLO) parton model \cite{Owens:1986mp} to calculate single inclusive hadron, di-hadron and $\gamma$-hadron spectra in proton-proton (p+p) and A+A collisions. Cross sections are factorized into parton distribution functions (PDF) of the proton or nuclei, partonic scattering cross sections, and parton fragmentation functions (FF). The PDF's of a free nucleon are given by the CT14 parameterization~\cite{Hou:2016nqm}. The FF's in vacuum are given by the Kniehl-Kramer-Potter parameterization \cite{Kniehl:2000fe}. 
The nuclear PDF's are from the EPPS16 \cite{Eskola:2016oht} parameterization with a model for the impact-parameter dependence \cite{Wang:1998ww,Hirano:2003pw,longpaper}. In nuclear collisions, medium-modified FF's are assumed to be given by the vacuum ones with reduced parton energy due to energy loss plus fragmentation of radiated gluons~\cite{Wang:2004yv,Zhang:2007ja,Zhang:2009rn}.

The radiative energy loss of a parton with momentum $p^\mu=(E,\vec p)$ within the higher-twist approach \cite{Guo:2000nz, Wang:2001ifa} is,
\begin{eqnarray}
\frac{\Delta{E}_a}{E} &=& \frac{2C_A\alpha_s}{\pi} \!\!\int_{\tau_0}^{\infty}\!\!\!\! d\tau \!\!\int \frac{dl_{\rm T}^2}{l_{\rm T}^2 (l_{\rm T}^2+\mu_D^2)}\!\!\int\!\! dz  \left[1+(1-z)^2\right] \nonumber\\
	&&\times \frac{p^\mu\cdot u_\mu}{E} \hat{q}_a(T(\tau)) \sin^2\left[\frac{l_{\rm T}^2(\tau-\tau_0)}{4z(1-z)E}\right],
\label{eq:deltaE}
\end{eqnarray}
where $\mu_D$ is the Debye mass, $C_A=3$, $\alpha_s$ the strong coupling constant, $l_{\rm T}$ the transverse momentum and $z$ the longitudinal momentum fraction of the radiated gluon. The time integral is along the jet path starting at time $\tau_0=0.6$ fm/$c$. The jet transport coefficient of a gluon $\hat q_A$ is $9/4$ times that of a quark $\hat q_F$. In this paper, the jet transport coefficient $\hat q(T)$ always refers to that of a quark. 
It is assumed to depend on the local temperature $T$ and  negligible in the hadron phase below $T_c=165$ MeV. The space-time profiles of $T$ and the four-velocity of the QGP fluid ($u^\mu$) are provided by the CLVisc 3+1 D hydrodynamic simulations~\cite{Pang:2012he,Pang:2014ipa,Pang:2018zzo} with
initial conditions given by the TRENTo model~\cite{Moreland:2014oya} averaged over 200 events for each centrality bin \footnote{An overall envelope function in the spatial rapidity is used to generalized the TRENTo initial condition at middle rapidity to a 3D distribution.}.
Parameters in the initial condition and the QGP transport coefficients are fitted to data on bulk hadron production. Uncertainties on the inferred $\hat q$ due to variations of these medium-related parameters are not the focus of this work 
and are not considered.

\noindent {\color{blue}\em Bayesian Inference with Information Field --}
A prior distribution with sufficient generality is critical for an unbiased extraction of physical parameters \cite{JETSCAPE:2020mzn}. However, the ``generality'' is a subtle issue for unknown functions. Priors according to an explicit parameterization of $\hat q(T)$ can introduce unnecessary long-range correlations between different regions of temperature. While peripheral or low energy collisions can only provide constraints at low $T$, $\hat q(T)$ at high $T$ can only be calibrated by data from central collisions at high colliding energies. One can invent parameterizations with higher flexibility \cite{Ke:2019jbh}, but it can become intractable for the inference of higher-dimensional functions.  We will develop an IF approach in this study to avoid this problem.

The IF approach views the unconditioned function as a random field. The Gaussian random field ($F$) is widely used, specified by a mean and a covariance function,
\begin{eqnarray}
      \langle F(x) \rangle &= \mu(x),~~
      \langle\delta F(x) \delta F(x') \rangle &= C(x,x^\prime),
\end{eqnarray}
where $\delta F(x) = F(x)-\mu(x)$ and higher cumulants vanish. Since $\hat q(T)$ in QGP is proportional to $T^3$, we will present the scaled quantity $F(x)=\ln(\hat q/T^3)$ as a random field in variable $x=\ln(T/{\rm GeV})$, which guarantees the positivity of $\hat{q}/T^3$ \footnote{Additional
constraints can be imposed by other pre-processing procedures.}.
For the scaled quantity $\hat q/T^3$, it is reasonable to assume a prior that has $\mu=\textrm{const}$. We assume that the correlation function takes the Gaussian form,
\begin{eqnarray}
C(x,x^\prime) = \sigma^2 \exp\left[-(x-x^\prime)^2/2l^2\right],
\end{eqnarray}
where $\sigma$ controls the variation of the random function with respect to the mean. The correlation length $l$ captures the essence of the IF approach. It states that the functional values at inputs $x$ and $x^\prime$ are decorrelated from each other for $|x-x^\prime|\gg l$. For the analysis of $\hat q(T)$, this translates to a condition that the high-temperature prior is unaffected by data that are only sensitive to $\hat q(T)$ at low temperature. This is the key to overcoming the problem of unnecessary long-range correlations that troubles the approach with explicit parameterizations.

\begin{figure}
    \centering
    \includegraphics[width=0.9\columnwidth]{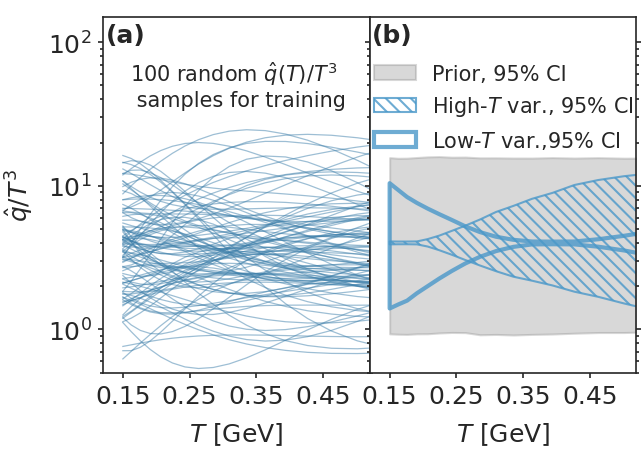}
    \caption{(a) 100 prior random functions for $\hat q/T^3$ from the IF approach. (b) Prior random functions for $\hat{q}/T^3$ at 95\% CI that are restricted to $\hat q/T^3=4\pm 0.1$ at $0.15<T<0.2$ GeV (High-$T$ var., hatched) or $0.3<T<0.4$ GeV (Low-$T$ var., unhatched) as compared to unconditioned prior (grey).}
    \label{fig:design}
\end{figure}

The IF prior only imposes continuity, differential properties, and local correlations on realizations of the random function. This is compatible with most physical quantities unless there is a discontinuity (e.g., at phase boundaries). The prior distribution in a compact form is
\begin{eqnarray}
P_0[F] =  \exp\left[\!-\frac{1}{2}\!\int\!\! dx dx^\prime \delta F(x) C^{-1}(x,x^\prime)\delta F(x^\prime)\right].
\end{eqnarray}
In the field theory language, the prior appears as the partition function of a free theory with the mean field $\mu(x)$  and the  propagator $C(x,x^\prime)$. The posterior distribution,
\begin{eqnarray}
P_1[F] &=& P_0[F] \exp\left\{-\ln \mathcal{L}({\rm Model}[F], {\rm Data})\right\},
\end{eqnarray}
will be determined by the likelihood function  $\mathcal{L}$  that depends on the pre-processed random function via the model and data comparison. The marginalized distribution of $y=F(x^*)$ can be expressed as the path integral $P(y) = \int [\mathcal{D}F] P_1[F] \delta(y-F[x^*])$.

In principle, the IF parameters $\mu, \sigma$ and $l$ should be treated as hyper-parameters and marginalized in the Bayesian analysis. As a first application of the IF approach, we fix their values to $\mu=\langle\ln \hat{q}/T^3\rangle =1.36$, $\sigma=0.7$ and $l=\ln(2)$. The values of $\mu$ and $\sigma$ are chosen so that the 95\% credible interval (CI) of the prior covers most of the range $0.8\lesssim\hat{q}/T^3 \lesssim 15$ from past analyses as shown (gray band) in Fig.~\ref{fig:design}(b). The choice of $l$ gives a smooth random function when $T$ varies within a factor of two while correlations beyond that are suppressed. Smaller values of $l$ can lead to short range oscillations in the random functions and  more samplings are required to reach the same accuracy in the final results~\cite{longpaper}.

\begin{figure}
\centering
\includegraphics[width=0.8\columnwidth]{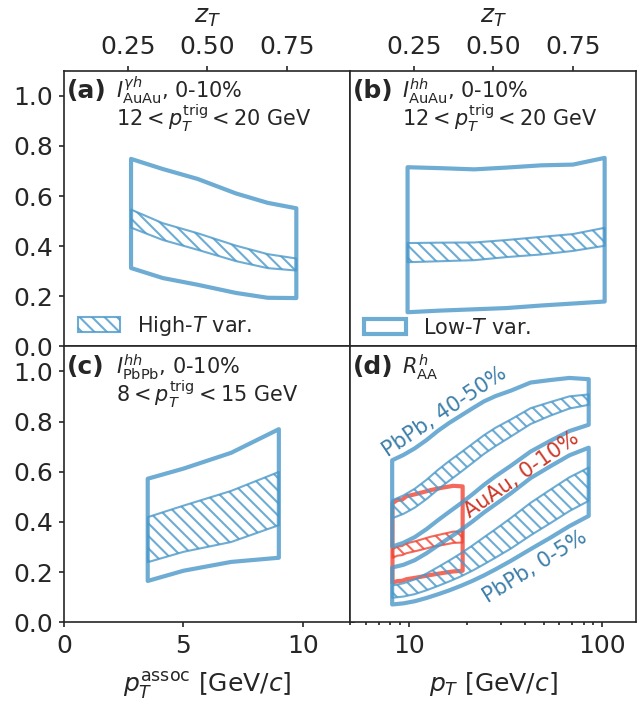}
    \caption{95\% variation of the ensemble predictions for (a) $I_{\rm AA}^{\gamma h}(z_{\rm T})$, (b) $I_{\rm AA}^{hh}(z_{\rm T})$ in 0-10\% Au+Au collisions at $\sqrt{s}=0.2$ TeV, (c) $I_{\rm AA}^{hh}(p_{\rm T}^{\rm assoc})$ in 0-10\% Pb+Pb collisions at $\sqrt{s}=2.76$ TeV, (d) $R_{\rm AA}^h(p_{\rm T})$ in 0-10\% Au+Au at $\sqrt{s}=0.2$ TeV (red), 0-5\% and 40-50\% Pb+Pb collisions at $\sqrt{s}=2.76$ TeV, using $\hat{q}/T^3$ sampled 
    from two sets of random functions, High-$T$ var. (hatched) and Low-$T$ var. (unhatched) as shown in Fig.~\ref{fig:design}(b).}
\label{fig:sensitivity}
\end{figure}

In our analysis, Gaussian Process (GP) emulators are used to speed up model predictions as in other studies (e.g., see Refs. \cite{Bernhard:2018hnz} and \cite{JETSCAPE:2020mzn} for detailed descriptions). We select 100 prior random function samples, shown in Fig.~\ref{fig:design}(a), according to the IF approach. With each $\hat{q}(T)$ sample, we use the NLO parton model to compute: 1) the nuclear suppression factor $R_{\rm AA}^h(p_{\rm T})$ for single inclusive hadrons defined as the ratio of normalized (by the number of binary collisions) $p_{\rm T}$ spectra in A+A and p+p collisions; 2) the nuclear modification factor $I_{\rm AA}^{hh}$ of di-hadron and 3) $I_{\rm AA}^{\gamma h}$ of $\gamma$-hadron correlation, defined as the ratio of hadron yield per trigger in A+A and p+p collisions as a function of the associate hadron $p_{\rm T}^{\rm assoc}$ or fractional momentum $z_{\rm T}=p_{\rm T}^{\rm assoc}/p_{\rm T}^{\rm trig}$. The 100 samples of $\hat{q}(T)$ and the corresponding model calculations of $R_{AA}^h$, $I_{\rm AA}^{hh}$, and $I_{\rm AA}^{\gamma h}$ for different centralities at both RHIC and LHC energies are used to train the GP emulators.

\noindent {\color{blue}\em Information field assisted sensitivity analysis --}
To demonstrate the sensitivity of different observables to the IF priors of $\hat q(T)$ in different temperature ranges, we create two sets of conditional random functions, as illustrated in Fig.~\ref{fig:design}(b).  In ``High-T var.'' prior, the random functions are constrained to $\hat{q}/T^3 = 4\pm 0.1$ at $0.15<T<0.2$ GeV but are allowed to vary in the high temperature region. The ``Low-T var.'' prior, on the other hand, is restricted to $\hat{q}/T^3 = 4\pm 0.1$ at $0.3<T<0.4$ GeV and can vary at low temperatures almost in the full range of prior values of $\hat{q}/T^3$. It is clear that the prior at high $T$ and low $T$ are indeed uncorrelated. The distance of decorrelation is controlled by $l$. The predicted observables with these two sets of priors are shown in Fig.~\ref{fig:sensitivity}. As expected, the sensitivities (variations) of $R_{\rm AA}^h$ and $I_{\rm AA}$ to high-$T$ $\hat{q}$ increase from RHIC to LHC energies, and from peripheral to central collisions. Furthermore, $I_{\rm AA}^{hh}$ and $I_{\rm AA}^{\gamma h}$ are slightly more sensitive to high-$T$ $\hat{q}$ than $R_{\rm AA}^h$ because of their different geometric bias on the initial jet production location in the transverse plane \cite{Zhang:2007ja,Zhang:2009rn}.

\begin{figure}
    \centering
    \includegraphics[width=0.9\columnwidth]{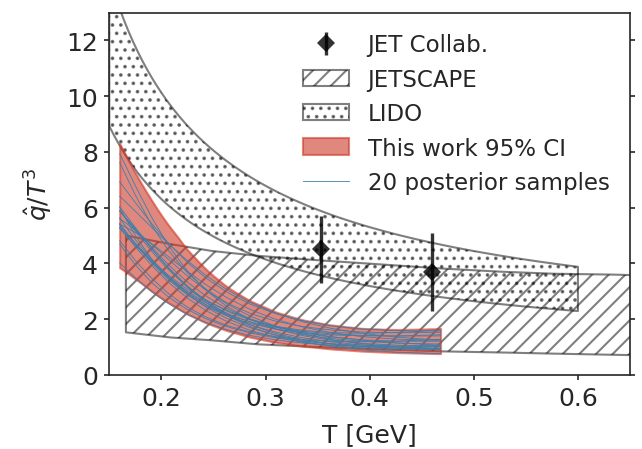}
    \caption{$\hat{q}/T^3$ from the global IF-based Bayesian analysis at 95\% CI (red) with 20 posterior samples (lines) as compared to results from JET Collaboration (symbols) \cite{JET:2013cls} and JETSCAPE (line-hatched) \cite{JETSCAPE:2021ehl} and LIDO analysis (dot-hatched) \cite{Ke:2020clc}.}
    \label{fig:qhat}
\end{figure}

\noindent {\color{blue}\em Global Bayesian inference of $\hat q(T)$--}
Using model emulators and all existing experimental data on $R_{\rm AA}^h$, $I^{hh}_{\rm AA}$ and $I^{\gamma h}_{\rm AA}$ at both RHIC and LHC energies, we have carried out the Bayesian inference of $\hat q(T)$.  Shown in Fig.~\ref{fig:qhat} is the final posterior  $\hat{q}/T^3$ (red band) in 95\% CI as compared to the fit by the JET Collaboration \cite{JET:2013cls} (symbols) and the JETSCAPE (95\% CI, line-hatched) \cite{JETSCAPE:2021ehl} and LIDO (95\% CI, dot-hatched) \cite{Ke:2020clc} Bayesian analyses. In addition, we also draw 20 random posterior samples (blue lines) to illustrate the distribution and correlation of the posterior $\hat q/T^3$. The IF approach can be extended in the future to include momentum and virtuality dependence of $\hat{q}$ as in the JETSCAPE analysis. With separate analyses in different regions of hadron's $p_{\rm T}$, we find only mild momentum dependence within $8<p_{\rm T}<100$ GeV/$c$~\cite{longpaper}. 

\begin{figure}
    \centering
    \includegraphics[width=0.9\columnwidth]{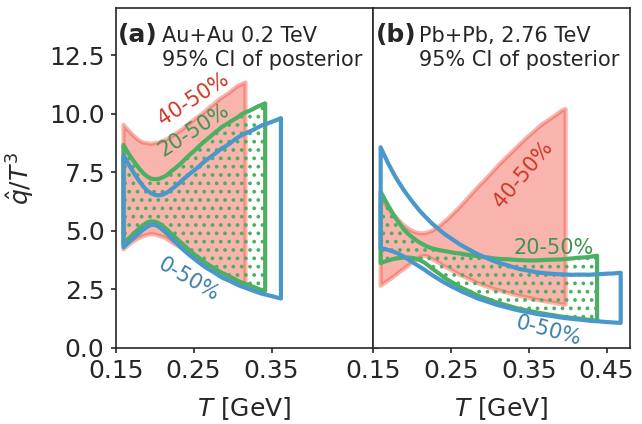}
    \caption{$\hat q/T^3$ as a function of temperature $T$ from IF-based Bayesian inference at 95\% CI using only data on $R_{\rm AA}^h(p_T)$ in (a) Au+Au at $\sqrt{s}=0.2$ TeV and (b) Pb+Pb collisions at $\sqrt{s}=2.76$ TeV with different ranges of centralities.}
    \label{fig:qhat2}
\end{figure}

The extracted $\hat q/T^3$ from the IF-based Bayesian inference is consistent with the JETSCAPE result \cite{JETSCAPE:2021ehl} at 95\% CI except at temperatures close to $T_c$ where a stronger temperature dependence arises from the combined constraint by experimental data in central and peripheral collisions at different colliding energies. The power of the combined constraint with the IF-based Bayesian inference is illustrated in Fig.~\ref{fig:qhat2} by the posterior $\hat q/T^3$ at 95\% CI using only $R_{\rm AA}^h$ data in (a) Au+Au at $\sqrt{s}=0.2$ TeV and (b) Pb+Pb collisions at $\sqrt{s}=2.76$ TeV with data from different range of centralities. Though data from peripheral collisions only provide meaningful constraints at lower $T$, $\hat q$ is progressively more constrained at higher $T$ as one includes data from more central collisions which probe higher $T$ regions of QGP.
We see the same trend as data from low to higher colliding energies are included. Because the IF approach strongly suppresses correlations between high and low-$T$ prior values of $\hat{q}$, it guarantees that constraints at higher $T$ are imposed only by combining data from more central collisions and at higher colliding energies. Including $I_{\rm AA}^{hh}$ and $I_{\rm AA}^{\gamma h}$ in the analysis does not significantly reduce the uncertainties because of the large experimental errors~\cite{longpaper}. This may be improved with more accurate data in the future.

\begin{figure}
    \centering
    \includegraphics[width=0.8\columnwidth]{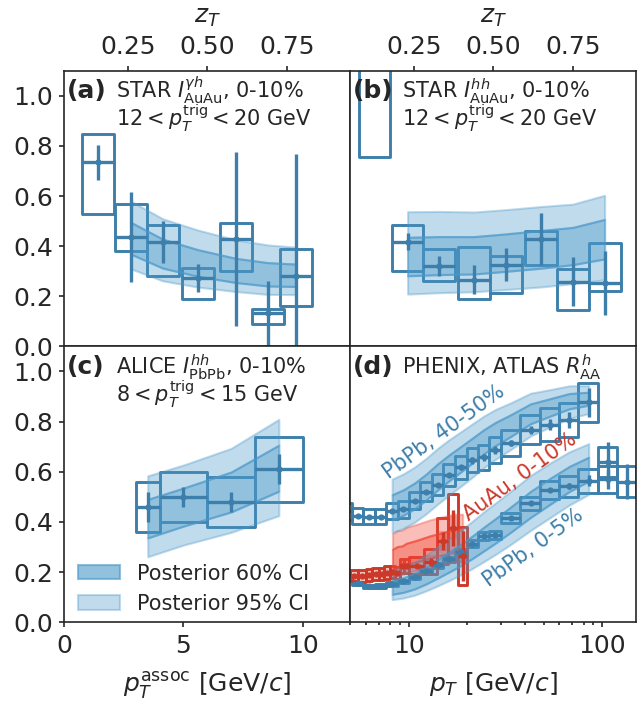}
    \caption{The same as Fig.~\ref{fig:sensitivity} except for posterior distributions of observables at 95\% (light blue) and 60\% CI (dark blue) from the model emulator with $\hat q/T^3$ given in Fig.~\ref{fig:qhat} as compared to a subset of experimental data~\cite{STAR:2016jdz,Aamodt:2011vg,Aad:2015wga,Adare:2008qa}.}
    \label{fig:all-obs}
\end{figure}

Finally, we compare in Fig.~\ref{fig:all-obs} the model emulator predictions with the globally Bayesian constrained $\hat q$ to a representative selection of experimental data on $R_{\rm AA}^h$, $I^{hh}_{\rm AA}$ and $I^{\gamma h}_{\rm AA}$ at both RHIC and LHC energies. A full comparison can be found in Ref.~\cite{longpaper}. Overall, the experimental data over a large range of $p_{\rm T}$, centralities and colliding energies are well reproduced by the ensemble predictions of the model emulator at 60\% and 95\% CI, indicating the global descriptive power of our analysis.

\noindent{\color{blue}\em Summary --}
We developed an Information Field (IF) approach to the Bayesian inference of unknown physical functions of system variables from experimental data. It represents a prior functional distribution that is free from long-range correlations between physical parameters in different regions of the variable space, which can be independently sensitive to different data sets. We applied this IF approach to the first global Bayesian inference of the jet transport coefficient $\hat q$ from combined experimental data on $R_{\rm AA}^h$, $I_{\rm AA}^{hh}$ and $I_{\rm AA}^{\gamma h}$ in heavy-ion collisions at both RHIC and LHC energies. We showed the IF approach can provide progressive constraints on $\hat q$ from low to high $T$ when experimental data in more central collisions and at higher colliding energies are included in the analysis. The extracted $\hat q/T^3$ is consistent with previous studies but exhibits a stronger temperature dependence. This IF approach can be extended to also include energy dependence of $\hat q$ and analyses of other properties of QGP such as bulk transport coefficients and equation of state.

\begin{acknowledgments}
We thank Chi Ding for providing the QGP hydro profiles. This work is supported by NSFC under Grant Nos. 11935007, 11861131009, 11890714 and 12075098, by  U.S. DOE under Contract No. DE-AC02-05CH11231, by US NSF under Grant OAC-2004571 within the X-SCAPE Collaboration, and by the LDRD Program at LANL. Computations are performed at NSC3 and NERSC.
\end{acknowledgments}

\bibliography{prc}

\begin{thebibliography}{72}%
\makeatletter
\providecommand \@ifxundefined [1]{%
 \@ifx{#1\undefined}
}%
\providecommand \@ifnum [1]{%
 \ifnum #1\expandafter \@firstoftwo
 \else \expandafter \@secondoftwo
 \fi
}%
\providecommand \@ifx [1]{%
 \ifx #1\expandafter \@firstoftwo
 \else \expandafter \@secondoftwo
 \fi
}%
\providecommand \natexlab [1]{#1}%
\providecommand \enquote  [1]{``#1''}%
\providecommand \bibnamefont  [1]{#1}%
\providecommand \bibfnamefont [1]{#1}%
\providecommand \citenamefont [1]{#1}%
\providecommand \href@noop [0]{\@secondoftwo}%
\providecommand \href [0]{\begingroup \@sanitize@url \@href}%
\providecommand \@href[1]{\@@startlink{#1}\@@href}%
\providecommand \@@href[1]{\endgroup#1\@@endlink}%
\providecommand \@sanitize@url [0]{\catcode `\\12\catcode `\$12\catcode
  `\&12\catcode `\#12\catcode `\^12\catcode `\_12\catcode `\%12\relax}%
\providecommand \@@startlink[1]{}%
\providecommand \@@endlink[0]{}%
\providecommand \url  [0]{\begingroup\@sanitize@url \@url }%
\providecommand \@url [1]{\endgroup\@href {#1}{\urlprefix }}%
\providecommand \urlprefix  [0]{URL }%
\providecommand \Eprint [0]{\href }%
\providecommand \doibase [0]{https://doi.org/}%
\providecommand \selectlanguage [0]{\@gobble}%
\providecommand \bibinfo  [0]{\@secondoftwo}%
\providecommand \bibfield  [0]{\@secondoftwo}%
\providecommand \translation [1]{[#1]}%
\providecommand \BibitemOpen [0]{}%
\providecommand \bibitemStop [0]{}%
\providecommand \bibitemNoStop [0]{.\EOS\space}%
\providecommand \EOS [0]{\spacefactor3000\relax}%
\providecommand \BibitemShut  [1]{\csname bibitem#1\endcsname}%
\let\auto@bib@innerbib\@empty
\bibitem [{\citenamefont {Pratt}\ \emph {et~al.}(2015)\citenamefont {Pratt},
  \citenamefont {Sangaline}, \citenamefont {Sorensen},\ and\ \citenamefont
  {Wang}}]{Pratt:2015zsa}%
  \BibitemOpen
  \bibfield  {author} {\bibinfo {author} {\bibfnamefont {S.}~\bibnamefont
  {Pratt}}, \bibinfo {author} {\bibfnamefont {E.}~\bibnamefont {Sangaline}},
  \bibinfo {author} {\bibfnamefont {P.}~\bibnamefont {Sorensen}},\ and\
  \bibinfo {author} {\bibfnamefont {H.}~\bibnamefont {Wang}},\ }\bibfield
  {title} {\bibinfo {title} {{Constraining the Eq. of State of Super-Hadronic
  Matter from Heavy-Ion Collisions}},\ }\href
  {https://doi.org/10.1103/PhysRevLett.114.202301} {\bibfield  {journal}
  {\bibinfo  {journal} {Phys. Rev. Lett.}\ }\textbf {\bibinfo {volume} {114}},\
  \bibinfo {pages} {202301} (\bibinfo {year} {2015})},\ \Eprint
  {https://arxiv.org/abs/1501.04042} {arXiv:1501.04042 [nucl-th]} \BibitemShut
  {NoStop}%
\bibitem [{\citenamefont {Bernhard}\ \emph {et~al.}(2015)\citenamefont
  {Bernhard}, \citenamefont {Marcy}, \citenamefont {Coleman-Smith},
  \citenamefont {Huzurbazar}, \citenamefont {Wolpert},\ and\ \citenamefont
  {Bass}}]{Bernhard:2015hxa}%
  \BibitemOpen
  \bibfield  {author} {\bibinfo {author} {\bibfnamefont {J.~E.}\ \bibnamefont
  {Bernhard}}, \bibinfo {author} {\bibfnamefont {P.~W.}\ \bibnamefont {Marcy}},
  \bibinfo {author} {\bibfnamefont {C.~E.}\ \bibnamefont {Coleman-Smith}},
  \bibinfo {author} {\bibfnamefont {S.}~\bibnamefont {Huzurbazar}}, \bibinfo
  {author} {\bibfnamefont {R.~L.}\ \bibnamefont {Wolpert}},\ and\ \bibinfo
  {author} {\bibfnamefont {S.~A.}\ \bibnamefont {Bass}},\ }\bibfield  {title}
  {\bibinfo {title} {{Quantifying properties of hot and dense QCD matter
  through systematic model-to-data comparison}},\ }\href
  {https://doi.org/10.1103/PhysRevC.91.054910} {\bibfield  {journal} {\bibinfo
  {journal} {Phys. Rev. C}\ }\textbf {\bibinfo {volume} {91}},\ \bibinfo
  {pages} {054910} (\bibinfo {year} {2015})},\ \Eprint
  {https://arxiv.org/abs/1502.00339} {arXiv:1502.00339 [nucl-th]} \BibitemShut
  {NoStop}%
\bibitem [{\citenamefont {Bernhard}\ \emph {et~al.}(2016)\citenamefont
  {Bernhard}, \citenamefont {Moreland}, \citenamefont {Bass}, \citenamefont
  {Liu},\ and\ \citenamefont {Heinz}}]{PhysRevC.94.024907}%
  \BibitemOpen
  \bibfield  {author} {\bibinfo {author} {\bibfnamefont {J.~E.}\ \bibnamefont
  {Bernhard}}, \bibinfo {author} {\bibfnamefont {J.~S.}\ \bibnamefont
  {Moreland}}, \bibinfo {author} {\bibfnamefont {S.~A.}\ \bibnamefont {Bass}},
  \bibinfo {author} {\bibfnamefont {J.}~\bibnamefont {Liu}},\ and\ \bibinfo
  {author} {\bibfnamefont {U.}~\bibnamefont {Heinz}},\ }\bibfield  {title}
  {\bibinfo {title} {Applying bayesian parameter estimation to relativistic
  heavy-ion collisions: Simultaneous characterization of the initial state and
  quark-gluon plasma medium},\ }\href
  {https://doi.org/10.1103/PhysRevC.94.024907} {\bibfield  {journal} {\bibinfo
  {journal} {Phys. Rev. C}\ }\textbf {\bibinfo {volume} {94}},\ \bibinfo
  {pages} {024907} (\bibinfo {year} {2016})}\BibitemShut {NoStop}%
\bibitem [{\citenamefont {Bernhard}\ \emph {et~al.}(2019)\citenamefont
  {Bernhard}, \citenamefont {Moreland},\ and\ \citenamefont
  {Bass}}]{Bernhard:2019bmu}%
  \BibitemOpen
  \bibfield  {author} {\bibinfo {author} {\bibfnamefont {J.~E.}\ \bibnamefont
  {Bernhard}}, \bibinfo {author} {\bibfnamefont {J.~S.}\ \bibnamefont
  {Moreland}},\ and\ \bibinfo {author} {\bibfnamefont {S.~A.}\ \bibnamefont
  {Bass}},\ }\bibfield  {title} {\bibinfo {title} {{Bayesian estimation of the
  specific shear and bulk viscosity of quark\textendash{}gluon plasma}},\
  }\href {https://doi.org/10.1038/s41567-019-0611-8} {\bibfield  {journal}
  {\bibinfo  {journal} {Nature Phys.}\ }\textbf {\bibinfo {volume} {15}},\
  \bibinfo {pages} {1113} (\bibinfo {year} {2019})}\BibitemShut {NoStop}%
\bibitem [{\citenamefont {Everett}\ \emph
  {et~al.}(2021{\natexlab{a}})\citenamefont {Everett} \emph
  {et~al.}}]{JETSCAPE:2020mzn}%
  \BibitemOpen
  \bibfield  {author} {\bibinfo {author} {\bibfnamefont {D.}~\bibnamefont
  {Everett}} \emph {et~al.} (\bibinfo {collaboration} {JETSCAPE}),\ }\bibfield
  {title} {\bibinfo {title} {{Multisystem Bayesian constraints on the transport
  coefficients of QCD matter}},\ }\href
  {https://doi.org/10.1103/PhysRevC.103.054904} {\bibfield  {journal} {\bibinfo
   {journal} {Phys. Rev. C}\ }\textbf {\bibinfo {volume} {103}},\ \bibinfo
  {pages} {054904} (\bibinfo {year} {2021}{\natexlab{a}})},\ \Eprint
  {https://arxiv.org/abs/2011.01430} {arXiv:2011.01430 [hep-ph]} \BibitemShut
  {NoStop}%
\bibitem [{\citenamefont {Everett}\ \emph
  {et~al.}(2021{\natexlab{b}})\citenamefont {Everett} \emph
  {et~al.}}]{JETSCAPE:2020shq}%
  \BibitemOpen
  \bibfield  {author} {\bibinfo {author} {\bibfnamefont {D.}~\bibnamefont
  {Everett}} \emph {et~al.} (\bibinfo {collaboration} {JETSCAPE}),\ }\bibfield
  {title} {\bibinfo {title} {{Phenomenological constraints on the transport
  properties of QCD matter with data-driven model averaging}},\ }\href
  {https://doi.org/10.1103/PhysRevLett.126.242301} {\bibfield  {journal}
  {\bibinfo  {journal} {Phys. Rev. Lett.}\ }\textbf {\bibinfo {volume} {126}},\
  \bibinfo {pages} {242301} (\bibinfo {year} {2021}{\natexlab{b}})},\ \Eprint
  {https://arxiv.org/abs/2010.03928} {arXiv:2010.03928 [hep-ph]} \BibitemShut
  {NoStop}%
\bibitem [{\citenamefont {Nijs}\ \emph
  {et~al.}(2021{\natexlab{a}})\citenamefont {Nijs}, \citenamefont {van~der
  Schee}, \citenamefont {G\"ursoy},\ and\ \citenamefont
  {Snellings}}]{Nijs:2020roc}%
  \BibitemOpen
  \bibfield  {author} {\bibinfo {author} {\bibfnamefont {G.}~\bibnamefont
  {Nijs}}, \bibinfo {author} {\bibfnamefont {W.}~\bibnamefont {van~der Schee}},
  \bibinfo {author} {\bibfnamefont {U.}~\bibnamefont {G\"ursoy}},\ and\
  \bibinfo {author} {\bibfnamefont {R.}~\bibnamefont {Snellings}},\ }\bibfield
  {title} {\bibinfo {title} {{Bayesian analysis of heavy ion collisions with
  the heavy ion computational framework Trajectum}},\ }\href
  {https://doi.org/10.1103/PhysRevC.103.054909} {\bibfield  {journal} {\bibinfo
   {journal} {Phys. Rev. C}\ }\textbf {\bibinfo {volume} {103}},\ \bibinfo
  {pages} {054909} (\bibinfo {year} {2021}{\natexlab{a}})},\ \Eprint
  {https://arxiv.org/abs/2010.15134} {arXiv:2010.15134 [nucl-th]} \BibitemShut
  {NoStop}%
\bibitem [{\citenamefont {Nijs}\ \emph
  {et~al.}(2021{\natexlab{b}})\citenamefont {Nijs}, \citenamefont {van~der
  Schee}, \citenamefont {G\"ursoy},\ and\ \citenamefont
  {Snellings}}]{Nijs:2020ors}%
  \BibitemOpen
  \bibfield  {author} {\bibinfo {author} {\bibfnamefont {G.}~\bibnamefont
  {Nijs}}, \bibinfo {author} {\bibfnamefont {W.}~\bibnamefont {van~der Schee}},
  \bibinfo {author} {\bibfnamefont {U.}~\bibnamefont {G\"ursoy}},\ and\
  \bibinfo {author} {\bibfnamefont {R.}~\bibnamefont {Snellings}},\ }\bibfield
  {title} {\bibinfo {title} {{Transverse Momentum Differential Global Analysis
  of Heavy-Ion Collisions}},\ }\href
  {https://doi.org/10.1103/PhysRevLett.126.202301} {\bibfield  {journal}
  {\bibinfo  {journal} {Phys. Rev. Lett.}\ }\textbf {\bibinfo {volume} {126}},\
  \bibinfo {pages} {202301} (\bibinfo {year} {2021}{\natexlab{b}})},\ \Eprint
  {https://arxiv.org/abs/2010.15130} {arXiv:2010.15130 [nucl-th]} \BibitemShut
  {NoStop}%
\bibitem [{\citenamefont {Parkkila}\ \emph
  {et~al.}(2021{\natexlab{a}})\citenamefont {Parkkila}, \citenamefont
  {Onnerstad},\ and\ \citenamefont {Kim}}]{Parkkila:2021tqq}%
  \BibitemOpen
  \bibfield  {author} {\bibinfo {author} {\bibfnamefont {J.~E.}\ \bibnamefont
  {Parkkila}}, \bibinfo {author} {\bibfnamefont {A.}~\bibnamefont
  {Onnerstad}},\ and\ \bibinfo {author} {\bibfnamefont {D.~J.}\ \bibnamefont
  {Kim}},\ }\bibfield  {title} {\bibinfo {title} {{Bayesian estimation of the
  specific shear and bulk viscosity of the quark-gluon plasma with additional
  flow harmonic observables}},\ }\href
  {https://doi.org/10.1103/PhysRevC.104.054904} {\bibfield  {journal} {\bibinfo
   {journal} {Phys. Rev. C}\ }\textbf {\bibinfo {volume} {104}},\ \bibinfo
  {pages} {054904} (\bibinfo {year} {2021}{\natexlab{a}})},\ \Eprint
  {https://arxiv.org/abs/2106.05019} {arXiv:2106.05019 [hep-ph]} \BibitemShut
  {NoStop}%
\bibitem [{\citenamefont {Parkkila}\ \emph
  {et~al.}(2021{\natexlab{b}})\citenamefont {Parkkila}, \citenamefont
  {Onnerstad}, \citenamefont {Taghavi}, \citenamefont {Mordasini},
  \citenamefont {Bilandzic},\ and\ \citenamefont {Kim}}]{Parkkila:2021yha}%
  \BibitemOpen
  \bibfield  {author} {\bibinfo {author} {\bibfnamefont {J.~E.}\ \bibnamefont
  {Parkkila}}, \bibinfo {author} {\bibfnamefont {A.}~\bibnamefont {Onnerstad}},
  \bibinfo {author} {\bibfnamefont {F.}~\bibnamefont {Taghavi}}, \bibinfo
  {author} {\bibfnamefont {C.}~\bibnamefont {Mordasini}}, \bibinfo {author}
  {\bibfnamefont {A.}~\bibnamefont {Bilandzic}},\ and\ \bibinfo {author}
  {\bibfnamefont {D.~J.}\ \bibnamefont {Kim}},\ }\bibfield  {title} {\bibinfo
  {title} {{New constraints for QCD matter from improved Bayesian parameter
  estimation in heavy-ion collisions at LHC}},\ }\href@noop {} {\  (\bibinfo
  {year} {2021}{\natexlab{b}})},\ \Eprint {https://arxiv.org/abs/2111.08145}
  {arXiv:2111.08145 [hep-ph]} \BibitemShut {NoStop}%
\bibitem [{\citenamefont {Xu}\ \emph {et~al.}(2018)\citenamefont {Xu},
  \citenamefont {Bernhard}, \citenamefont {Bass}, \citenamefont {Nahrgang},\
  and\ \citenamefont {Cao}}]{Xu:2017obm}%
  \BibitemOpen
  \bibfield  {author} {\bibinfo {author} {\bibfnamefont {Y.}~\bibnamefont
  {Xu}}, \bibinfo {author} {\bibfnamefont {J.~E.}\ \bibnamefont {Bernhard}},
  \bibinfo {author} {\bibfnamefont {S.~A.}\ \bibnamefont {Bass}}, \bibinfo
  {author} {\bibfnamefont {M.}~\bibnamefont {Nahrgang}},\ and\ \bibinfo
  {author} {\bibfnamefont {S.}~\bibnamefont {Cao}},\ }\bibfield  {title}
  {\bibinfo {title} {{Data-driven analysis for the temperature and momentum
  dependence of the heavy-quark diffusion coefficient in relativistic heavy-ion
  collisions}},\ }\href {https://doi.org/10.1103/PhysRevC.97.014907} {\bibfield
   {journal} {\bibinfo  {journal} {Phys. Rev. C}\ }\textbf {\bibinfo {volume}
  {97}},\ \bibinfo {pages} {014907} (\bibinfo {year} {2018})},\ \Eprint
  {https://arxiv.org/abs/1710.00807} {arXiv:1710.00807 [nucl-th]} \BibitemShut
  {NoStop}%
\bibitem [{\citenamefont {Ke}\ \emph {et~al.}(2018)\citenamefont {Ke},
  \citenamefont {Xu},\ and\ \citenamefont {Bass}}]{PhysRevC.98.064901}%
  \BibitemOpen
  \bibfield  {author} {\bibinfo {author} {\bibfnamefont {W.}~\bibnamefont
  {Ke}}, \bibinfo {author} {\bibfnamefont {Y.}~\bibnamefont {Xu}},\ and\
  \bibinfo {author} {\bibfnamefont {S.~A.}\ \bibnamefont {Bass}},\ }\bibfield
  {title} {\bibinfo {title} {Linearized boltzmann-langevin model for heavy
  quark transport in hot and dense qcd matter},\ }\href
  {https://doi.org/10.1103/PhysRevC.98.064901} {\bibfield  {journal} {\bibinfo
  {journal} {Phys. Rev. C}\ }\textbf {\bibinfo {volume} {98}},\ \bibinfo
  {pages} {064901} (\bibinfo {year} {2018})}\BibitemShut {NoStop}%
\bibitem [{\citenamefont {Liu}\ \emph {et~al.}(2022)\citenamefont {Liu},
  \citenamefont {Xing}, \citenamefont {Wu}, \citenamefont {Qin}, \citenamefont
  {Cao},\ and\ \citenamefont {Wang}}]{Liu:2021dpm}%
  \BibitemOpen
  \bibfield  {author} {\bibinfo {author} {\bibfnamefont {F.-L.}\ \bibnamefont
  {Liu}}, \bibinfo {author} {\bibfnamefont {W.-J.}\ \bibnamefont {Xing}},
  \bibinfo {author} {\bibfnamefont {X.-Y.}\ \bibnamefont {Wu}}, \bibinfo
  {author} {\bibfnamefont {G.-Y.}\ \bibnamefont {Qin}}, \bibinfo {author}
  {\bibfnamefont {S.}~\bibnamefont {Cao}},\ and\ \bibinfo {author}
  {\bibfnamefont {X.-N.}\ \bibnamefont {Wang}},\ }\bibfield  {title} {\bibinfo
  {title} {{QLBT: a linear Boltzmann transport model for heavy quarks in a
  quark-gluon plasma of quasi-particles}},\ }\href
  {https://doi.org/10.1140/epjc/s10052-022-10308-x} {\bibfield  {journal}
  {\bibinfo  {journal} {Eur. Phys. J. C}\ }\textbf {\bibinfo {volume} {82}},\
  \bibinfo {pages} {350} (\bibinfo {year} {2022})},\ \Eprint
  {https://arxiv.org/abs/2107.11713} {arXiv:2107.11713 [hep-ph]} \BibitemShut
  {NoStop}%
\bibitem [{\citenamefont {Ke}\ and\ \citenamefont {Wang}(2021)}]{Ke:2020clc}%
  \BibitemOpen
  \bibfield  {author} {\bibinfo {author} {\bibfnamefont {W.}~\bibnamefont
  {Ke}}\ and\ \bibinfo {author} {\bibfnamefont {X.-N.}\ \bibnamefont {Wang}},\
  }\bibfield  {title} {\bibinfo {title} {{QGP modification to single inclusive
  jets in a calibrated transport model}},\ }\href
  {https://doi.org/10.1007/JHEP05(2021)041} {\bibfield  {journal} {\bibinfo
  {journal} {JHEP}\ }\textbf {\bibinfo {volume} {05}},\ \bibinfo {pages}
  {041}},\ \Eprint {https://arxiv.org/abs/2010.13680} {arXiv:2010.13680
  [hep-ph]} \BibitemShut {NoStop}%
\bibitem [{\citenamefont {Cao}\ \emph {et~al.}(2021)\citenamefont {Cao} \emph
  {et~al.}}]{JETSCAPE:2021ehl}%
  \BibitemOpen
  \bibfield  {author} {\bibinfo {author} {\bibfnamefont {S.}~\bibnamefont
  {Cao}} \emph {et~al.} (\bibinfo {collaboration} {JETSCAPE}),\ }\bibfield
  {title} {\bibinfo {title} {{Determining the jet transport coefficient
  $\hat{q}$ from inclusive hadron suppression measurements using Bayesian
  parameter estimation}},\ }\href {https://doi.org/10.1103/PhysRevC.104.024905}
  {\bibfield  {journal} {\bibinfo  {journal} {Phys. Rev. C}\ }\textbf {\bibinfo
  {volume} {104}},\ \bibinfo {pages} {024905} (\bibinfo {year} {2021})},\
  \Eprint {https://arxiv.org/abs/2102.11337} {arXiv:2102.11337 [nucl-th]}
  \BibitemShut {NoStop}%
\bibitem [{\citenamefont {Casalderrey-Solana}\ and\ \citenamefont
  {Wang}(2008)}]{Casalderrey-Solana:2007xns}%
  \BibitemOpen
  \bibfield  {author} {\bibinfo {author} {\bibfnamefont {J.}~\bibnamefont
  {Casalderrey-Solana}}\ and\ \bibinfo {author} {\bibfnamefont {X.-N.}\
  \bibnamefont {Wang}},\ }\bibfield  {title} {\bibinfo {title} {{Energy
  dependence of jet transport parameter and parton saturation in quark-gluon
  plasma}},\ }\href {https://doi.org/10.1103/PhysRevC.77.024902} {\bibfield
  {journal} {\bibinfo  {journal} {Phys. Rev. C}\ }\textbf {\bibinfo {volume}
  {77}},\ \bibinfo {pages} {024902} (\bibinfo {year} {2008})},\ \Eprint
  {https://arxiv.org/abs/0705.1352} {arXiv:0705.1352 [hep-ph]} \BibitemShut
  {NoStop}%
\bibitem [{\citenamefont {He}\ \emph {et~al.}(2015)\citenamefont {He},
  \citenamefont {Luo}, \citenamefont {Wang},\ and\ \citenamefont
  {Zhu}}]{He:2015pra}%
  \BibitemOpen
  \bibfield  {author} {\bibinfo {author} {\bibfnamefont {Y.}~\bibnamefont
  {He}}, \bibinfo {author} {\bibfnamefont {T.}~\bibnamefont {Luo}}, \bibinfo
  {author} {\bibfnamefont {X.-N.}\ \bibnamefont {Wang}},\ and\ \bibinfo
  {author} {\bibfnamefont {Y.}~\bibnamefont {Zhu}},\ }\bibfield  {title}
  {\bibinfo {title} {{Linear Boltzmann Transport for Jet Propagation in the
  Quark-Gluon Plasma: Elastic Processes and Medium Recoil}},\ }\href
  {https://doi.org/10.1103/PhysRevC.91.054908} {\bibfield  {journal} {\bibinfo
  {journal} {Phys. Rev. C}\ }\textbf {\bibinfo {volume} {91}},\ \bibinfo
  {pages} {054908} (\bibinfo {year} {2015})},\ \bibinfo {note} {[Erratum:
  Phys.Rev.C 97, 019902 (2018)]},\ \Eprint {https://arxiv.org/abs/1503.03313}
  {arXiv:1503.03313 [nucl-th]} \BibitemShut {NoStop}%
\bibitem [{\citenamefont {Lin}\ \emph {et~al.}(2018)\citenamefont {Lin},
  \citenamefont {Melnitchouk}, \citenamefont {Prokudin}, \citenamefont {Sato},\
  and\ \citenamefont {Shows}}]{PhysRevLett.120.152502}%
  \BibitemOpen
  \bibfield  {author} {\bibinfo {author} {\bibfnamefont {H.-W.}\ \bibnamefont
  {Lin}}, \bibinfo {author} {\bibfnamefont {W.}~\bibnamefont {Melnitchouk}},
  \bibinfo {author} {\bibfnamefont {A.}~\bibnamefont {Prokudin}}, \bibinfo
  {author} {\bibfnamefont {N.}~\bibnamefont {Sato}},\ and\ \bibinfo {author}
  {\bibfnamefont {H.}~\bibnamefont {Shows}} (\bibinfo {collaboration}
  {Jefferson Lab Angular Momentum (JAM) Collaboration}),\ }\bibfield  {title}
  {\bibinfo {title} {First monte carlo global analysis of nucleon transversity
  with lattice qcd constraints},\ }\href
  {https://doi.org/10.1103/PhysRevLett.120.152502} {\bibfield  {journal}
  {\bibinfo  {journal} {Phys. Rev. Lett.}\ }\textbf {\bibinfo {volume} {120}},\
  \bibinfo {pages} {152502} (\bibinfo {year} {2018})}\BibitemShut {NoStop}%
\bibitem [{\citenamefont {Del~Debbio}\ \emph {et~al.}(2022)\citenamefont
  {Del~Debbio}, \citenamefont {Giani},\ and\ \citenamefont
  {Wilson}}]{DelDebbio:2021whr}%
  \BibitemOpen
  \bibfield  {author} {\bibinfo {author} {\bibfnamefont {L.}~\bibnamefont
  {Del~Debbio}}, \bibinfo {author} {\bibfnamefont {T.}~\bibnamefont {Giani}},\
  and\ \bibinfo {author} {\bibfnamefont {M.}~\bibnamefont {Wilson}},\
  }\bibfield  {title} {\bibinfo {title} {{Bayesian approach to inverse
  problems: an application to NNPDF closure testing}},\ }\href
  {https://doi.org/10.1140/epjc/s10052-022-10297-x} {\bibfield  {journal}
  {\bibinfo  {journal} {Eur. Phys. J. C}\ }\textbf {\bibinfo {volume} {82}},\
  \bibinfo {pages} {330} (\bibinfo {year} {2022})},\ \Eprint
  {https://arxiv.org/abs/2111.05787} {arXiv:2111.05787 [hep-ph]} \BibitemShut
  {NoStop}%
\bibitem [{\citenamefont {Bialek}\ \emph {et~al.}(1996)\citenamefont {Bialek},
  \citenamefont {Callan},\ and\ \citenamefont {Strong}}]{PhysRevLett.77.4693}%
  \BibitemOpen
  \bibfield  {author} {\bibinfo {author} {\bibfnamefont {W.}~\bibnamefont
  {Bialek}}, \bibinfo {author} {\bibfnamefont {C.~G.}\ \bibnamefont {Callan}},\
  and\ \bibinfo {author} {\bibfnamefont {S.~P.}\ \bibnamefont {Strong}},\
  }\bibfield  {title} {\bibinfo {title} {Field theories for learning
  probability distributions},\ }\href
  {https://doi.org/10.1103/PhysRevLett.77.4693} {\bibfield  {journal} {\bibinfo
   {journal} {Phys. Rev. Lett.}\ }\textbf {\bibinfo {volume} {77}},\ \bibinfo
  {pages} {4693} (\bibinfo {year} {1996})}\BibitemShut {NoStop}%
\bibitem [{\citenamefont {Ensslin}(2013)}]{Ensslin:2013ji}%
  \BibitemOpen
  \bibfield  {author} {\bibinfo {author} {\bibfnamefont {T.}~\bibnamefont
  {Ensslin}},\ }\bibfield  {title} {\bibinfo {title} {{Information field
  theory}},\ }\href {https://doi.org/10.1063/1.4819999} {\bibfield  {journal}
  {\bibinfo  {journal} {AIP Conf. Proc.}\ }\textbf {\bibinfo {volume} {1553}},\
  \bibinfo {pages} {184} (\bibinfo {year} {2013})},\ \Eprint
  {https://arxiv.org/abs/1301.2556} {arXiv:1301.2556 [astro-ph.IM]}
  \BibitemShut {NoStop}%
\bibitem [{\citenamefont
  {Lemm}(1999)}]{https://doi.org/10.48550/arxiv.physics/9912005}%
  \BibitemOpen
  \bibfield  {author} {\bibinfo {author} {\bibfnamefont {J.~C.}\ \bibnamefont
  {Lemm}},\ }\href {https://doi.org/10.48550/ARXIV.PHYSICS/9912005} {\bibinfo
  {title} {Bayesian field theory: Nonparametric approaches to density
  estimation, regression, classification, and inverse quantum problems}}
  (\bibinfo {year} {1999})\BibitemShut {NoStop}%
\bibitem [{\citenamefont {Gyulassy}\ and\ \citenamefont
  {Plumer}(1990)}]{Gyulassy:1990ye}%
  \BibitemOpen
  \bibfield  {author} {\bibinfo {author} {\bibfnamefont {M.}~\bibnamefont
  {Gyulassy}}\ and\ \bibinfo {author} {\bibfnamefont {M.}~\bibnamefont
  {Plumer}},\ }\bibfield  {title} {\bibinfo {title} {{Jet Quenching in Dense
  Matter}},\ }\href {https://doi.org/10.1016/0370-2693(90)91409-5} {\bibfield
  {journal} {\bibinfo  {journal} {Phys. Lett. B}\ }\textbf {\bibinfo {volume}
  {243}},\ \bibinfo {pages} {432} (\bibinfo {year} {1990})}\BibitemShut
  {NoStop}%
\bibitem [{\citenamefont {Wang}\ and\ \citenamefont
  {Gyulassy}(1992)}]{Wang:1991xy}%
  \BibitemOpen
  \bibfield  {author} {\bibinfo {author} {\bibfnamefont {X.-N.}\ \bibnamefont
  {Wang}}\ and\ \bibinfo {author} {\bibfnamefont {M.}~\bibnamefont
  {Gyulassy}},\ }\bibfield  {title} {\bibinfo {title} {{Gluon shadowing and jet
  quenching in A + A collisions at s**(1/2) = 200-GeV}},\ }\href
  {https://doi.org/10.1103/PhysRevLett.68.1480} {\bibfield  {journal} {\bibinfo
   {journal} {Phys. Rev. Lett.}\ }\textbf {\bibinfo {volume} {68}},\ \bibinfo
  {pages} {1480} (\bibinfo {year} {1992})}\BibitemShut {NoStop}%
\bibitem [{\citenamefont {Wang}(1998)}]{Wang:1998bha}%
  \BibitemOpen
  \bibfield  {author} {\bibinfo {author} {\bibfnamefont {X.-N.}\ \bibnamefont
  {Wang}},\ }\bibfield  {title} {\bibinfo {title} {{Effect of jet quenching on
  high $p_{T}$ hadron spectra in high-energy nuclear collisions}},\ }\href
  {https://doi.org/10.1103/PhysRevC.58.2321} {\bibfield  {journal} {\bibinfo
  {journal} {Phys. Rev. C}\ }\textbf {\bibinfo {volume} {58}},\ \bibinfo
  {pages} {2321} (\bibinfo {year} {1998})},\ \Eprint
  {https://arxiv.org/abs/hep-ph/9804357} {arXiv:hep-ph/9804357} \BibitemShut
  {NoStop}%
\bibitem [{\citenamefont {Majumder}\ and\ \citenamefont
  {Van~Leeuwen}(2011)}]{Majumder:2010qh}%
  \BibitemOpen
  \bibfield  {author} {\bibinfo {author} {\bibfnamefont {A.}~\bibnamefont
  {Majumder}}\ and\ \bibinfo {author} {\bibfnamefont {M.}~\bibnamefont
  {Van~Leeuwen}},\ }\bibfield  {title} {\bibinfo {title} {{The Theory and
  Phenomenology of Perturbative QCD Based Jet Quenching}},\ }\href
  {https://doi.org/10.1016/j.ppnp.2010.09.001} {\bibfield  {journal} {\bibinfo
  {journal} {Prog. Part. Nucl. Phys.}\ }\textbf {\bibinfo {volume} {66}},\
  \bibinfo {pages} {41} (\bibinfo {year} {2011})},\ \Eprint
  {https://arxiv.org/abs/1002.2206} {arXiv:1002.2206 [hep-ph]} \BibitemShut
  {NoStop}%
\bibitem [{\citenamefont {Qin}\ and\ \citenamefont {Wang}(2015)}]{Qin:2015srf}%
  \BibitemOpen
  \bibfield  {author} {\bibinfo {author} {\bibfnamefont {G.-Y.}\ \bibnamefont
  {Qin}}\ and\ \bibinfo {author} {\bibfnamefont {X.-N.}\ \bibnamefont {Wang}},\
  }\bibfield  {title} {\bibinfo {title} {{Jet quenching in high-energy
  heavy-ion collisions}},\ }\href {https://doi.org/10.1142/S0218301315300143}
  {\bibfield  {journal} {\bibinfo  {journal} {Int. J. Mod. Phys. E}\ }\textbf
  {\bibinfo {volume} {24}},\ \bibinfo {pages} {1530014} (\bibinfo {year}
  {2015})},\ \Eprint {https://arxiv.org/abs/1511.00790} {arXiv:1511.00790
  [hep-ph]} \BibitemShut {NoStop}%
\bibitem [{\citenamefont {Baier}\ \emph
  {et~al.}(1997{\natexlab{a}})\citenamefont {Baier}, \citenamefont
  {Dokshitzer}, \citenamefont {Mueller}, \citenamefont {Peigne},\ and\
  \citenamefont {Schiff}}]{Baier:1996kr}%
  \BibitemOpen
  \bibfield  {author} {\bibinfo {author} {\bibfnamefont {R.}~\bibnamefont
  {Baier}}, \bibinfo {author} {\bibfnamefont {Y.~L.}\ \bibnamefont
  {Dokshitzer}}, \bibinfo {author} {\bibfnamefont {A.~H.}\ \bibnamefont
  {Mueller}}, \bibinfo {author} {\bibfnamefont {S.}~\bibnamefont {Peigne}},\
  and\ \bibinfo {author} {\bibfnamefont {D.}~\bibnamefont {Schiff}},\
  }\bibfield  {title} {\bibinfo {title} {{Radiative energy loss of high-energy
  quarks and gluons in a finite volume quark - gluon plasma}},\ }\href
  {https://doi.org/10.1016/S0550-3213(96)00553-6} {\bibfield  {journal}
  {\bibinfo  {journal} {Nucl. Phys. B}\ }\textbf {\bibinfo {volume} {483}},\
  \bibinfo {pages} {291} (\bibinfo {year} {1997}{\natexlab{a}})},\ \Eprint
  {https://arxiv.org/abs/hep-ph/9607355} {arXiv:hep-ph/9607355} \BibitemShut
  {NoStop}%
\bibitem [{\citenamefont {Baier}\ \emph
  {et~al.}(1997{\natexlab{b}})\citenamefont {Baier}, \citenamefont
  {Dokshitzer}, \citenamefont {Mueller}, \citenamefont {Peigne},\ and\
  \citenamefont {Schiff}}]{Baier:1996sk}%
  \BibitemOpen
  \bibfield  {author} {\bibinfo {author} {\bibfnamefont {R.}~\bibnamefont
  {Baier}}, \bibinfo {author} {\bibfnamefont {Y.~L.}\ \bibnamefont
  {Dokshitzer}}, \bibinfo {author} {\bibfnamefont {A.~H.}\ \bibnamefont
  {Mueller}}, \bibinfo {author} {\bibfnamefont {S.}~\bibnamefont {Peigne}},\
  and\ \bibinfo {author} {\bibfnamefont {D.}~\bibnamefont {Schiff}},\
  }\bibfield  {title} {\bibinfo {title} {{Radiative energy loss and p(T)
  broadening of high-energy partons in nuclei}},\ }\href
  {https://doi.org/10.1016/S0550-3213(96)00581-0} {\bibfield  {journal}
  {\bibinfo  {journal} {Nucl. Phys. B}\ }\textbf {\bibinfo {volume} {484}},\
  \bibinfo {pages} {265} (\bibinfo {year} {1997}{\natexlab{b}})},\ \Eprint
  {https://arxiv.org/abs/hep-ph/9608322} {arXiv:hep-ph/9608322} \BibitemShut
  {NoStop}%
\bibitem [{\citenamefont {Zakharov}(1996)}]{Zakharov:1996fv}%
  \BibitemOpen
  \bibfield  {author} {\bibinfo {author} {\bibfnamefont {B.~G.}\ \bibnamefont
  {Zakharov}},\ }\bibfield  {title} {\bibinfo {title} {{Fully quantum treatment
  of the Landau-Pomeranchuk-Migdal effect in QED and QCD}},\ }\href
  {https://doi.org/10.1134/1.567126} {\bibfield  {journal} {\bibinfo  {journal}
  {JETP Lett.}\ }\textbf {\bibinfo {volume} {63}},\ \bibinfo {pages} {952}
  (\bibinfo {year} {1996})},\ \Eprint {https://arxiv.org/abs/hep-ph/9607440}
  {arXiv:hep-ph/9607440} \BibitemShut {NoStop}%
\bibitem [{\citenamefont {Wiedemann}(2000)}]{Wiedemann:2000za}%
  \BibitemOpen
  \bibfield  {author} {\bibinfo {author} {\bibfnamefont {U.~A.}\ \bibnamefont
  {Wiedemann}},\ }\bibfield  {title} {\bibinfo {title} {{Gluon radiation off
  hard quarks in a nuclear environment: Opacity expansion}},\ }\href
  {https://doi.org/10.1016/S0550-3213(00)00457-0} {\bibfield  {journal}
  {\bibinfo  {journal} {Nucl. Phys. B}\ }\textbf {\bibinfo {volume} {588}},\
  \bibinfo {pages} {303} (\bibinfo {year} {2000})},\ \Eprint
  {https://arxiv.org/abs/hep-ph/0005129} {arXiv:hep-ph/0005129} \BibitemShut
  {NoStop}%
\bibitem [{\citenamefont {Guo}\ and\ \citenamefont {Wang}(2000)}]{Guo:2000nz}%
  \BibitemOpen
  \bibfield  {author} {\bibinfo {author} {\bibfnamefont {X.-F.}\ \bibnamefont
  {Guo}}\ and\ \bibinfo {author} {\bibfnamefont {X.-N.}\ \bibnamefont {Wang}},\
  }\bibfield  {title} {\bibinfo {title} {{Multiple scattering, parton energy
  loss and modified fragmentation functions in deeply inelastic e A
  scattering}},\ }\href {https://doi.org/10.1103/PhysRevLett.85.3591}
  {\bibfield  {journal} {\bibinfo  {journal} {Phys. Rev. Lett.}\ }\textbf
  {\bibinfo {volume} {85}},\ \bibinfo {pages} {3591} (\bibinfo {year}
  {2000})},\ \Eprint {https://arxiv.org/abs/hep-ph/0005044}
  {arXiv:hep-ph/0005044} \BibitemShut {NoStop}%
\bibitem [{\citenamefont {Wang}\ and\ \citenamefont
  {Guo}(2001)}]{Wang:2001ifa}%
  \BibitemOpen
  \bibfield  {author} {\bibinfo {author} {\bibfnamefont {X.-N.}\ \bibnamefont
  {Wang}}\ and\ \bibinfo {author} {\bibfnamefont {X.-F.}\ \bibnamefont {Guo}},\
  }\bibfield  {title} {\bibinfo {title} {{Multiple parton scattering in nuclei:
  Parton energy loss}},\ }\href {https://doi.org/10.1016/S0375-9474(01)01130-7}
  {\bibfield  {journal} {\bibinfo  {journal} {Nucl. Phys. A}\ }\textbf
  {\bibinfo {volume} {696}},\ \bibinfo {pages} {788} (\bibinfo {year}
  {2001})},\ \Eprint {https://arxiv.org/abs/hep-ph/0102230}
  {arXiv:hep-ph/0102230} \BibitemShut {NoStop}%
\bibitem [{\citenamefont {Majumder}(2012)}]{Majumder:2009ge}%
  \BibitemOpen
  \bibfield  {author} {\bibinfo {author} {\bibfnamefont {A.}~\bibnamefont
  {Majumder}},\ }\bibfield  {title} {\bibinfo {title} {{Hard collinear gluon
  radiation and multiple scattering in a medium}},\ }\href
  {https://doi.org/10.1103/PhysRevD.85.014023} {\bibfield  {journal} {\bibinfo
  {journal} {Phys. Rev. D}\ }\textbf {\bibinfo {volume} {85}},\ \bibinfo
  {pages} {014023} (\bibinfo {year} {2012})},\ \Eprint
  {https://arxiv.org/abs/0912.2987} {arXiv:0912.2987 [nucl-th]} \BibitemShut
  {NoStop}%
\bibitem [{\citenamefont {Burke}\ \emph {et~al.}(2014)\citenamefont {Burke}
  \emph {et~al.}}]{JET:2013cls}%
  \BibitemOpen
  \bibfield  {author} {\bibinfo {author} {\bibfnamefont {K.~M.}\ \bibnamefont
  {Burke}} \emph {et~al.} (\bibinfo {collaboration} {JET}),\ }\bibfield
  {title} {\bibinfo {title} {{Extracting the jet transport coefficient from jet
  quenching in high-energy heavy-ion collisions}},\ }\href
  {https://doi.org/10.1103/PhysRevC.90.014909} {\bibfield  {journal} {\bibinfo
  {journal} {Phys. Rev. C}\ }\textbf {\bibinfo {volume} {90}},\ \bibinfo
  {pages} {014909} (\bibinfo {year} {2014})},\ \Eprint
  {https://arxiv.org/abs/1312.5003} {arXiv:1312.5003 [nucl-th]} \BibitemShut
  {NoStop}%
\bibitem [{\citenamefont {Liao}\ and\ \citenamefont
  {Shuryak}(2009)}]{Liao:2008dk}%
  \BibitemOpen
  \bibfield  {author} {\bibinfo {author} {\bibfnamefont {J.}~\bibnamefont
  {Liao}}\ and\ \bibinfo {author} {\bibfnamefont {E.}~\bibnamefont {Shuryak}},\
  }\bibfield  {title} {\bibinfo {title} {{Angular Dependence of Jet Quenching
  Indicates Its Strong Enhancement Near the QCD Phase Transition}},\ }\href
  {https://doi.org/10.1103/PhysRevLett.102.202302} {\bibfield  {journal}
  {\bibinfo  {journal} {Phys. Rev. Lett.}\ }\textbf {\bibinfo {volume} {102}},\
  \bibinfo {pages} {202302} (\bibinfo {year} {2009})},\ \Eprint
  {https://arxiv.org/abs/0810.4116} {arXiv:0810.4116 [nucl-th]} \BibitemShut
  {NoStop}%
\bibitem [{\citenamefont {Xu}\ \emph {et~al.}(2015)\citenamefont {Xu},
  \citenamefont {Liao},\ and\ \citenamefont {Gyulassy}}]{Xu:2014tda}%
  \BibitemOpen
  \bibfield  {author} {\bibinfo {author} {\bibfnamefont {J.}~\bibnamefont
  {Xu}}, \bibinfo {author} {\bibfnamefont {J.}~\bibnamefont {Liao}},\ and\
  \bibinfo {author} {\bibfnamefont {M.}~\bibnamefont {Gyulassy}},\ }\bibfield
  {title} {\bibinfo {title} {{Consistency of Perfect Fluidity and Jet Quenching
  in semi-Quark-Gluon Monopole Plasmas}},\ }\href
  {https://doi.org/10.1088/0256-307X/32/9/092501} {\bibfield  {journal}
  {\bibinfo  {journal} {Chin. Phys. Lett.}\ }\textbf {\bibinfo {volume} {32}},\
  \bibinfo {pages} {092501} (\bibinfo {year} {2015})},\ \Eprint
  {https://arxiv.org/abs/1411.3673} {arXiv:1411.3673 [hep-ph]} \BibitemShut
  {NoStop}%
\bibitem [{\citenamefont {Adare}\ \emph {et~al.}(2008)\citenamefont {Adare}
  \emph {et~al.}}]{Adare:2008qa}%
  \BibitemOpen
  \bibfield  {author} {\bibinfo {author} {\bibfnamefont {A.}~\bibnamefont
  {Adare}} \emph {et~al.} (\bibinfo {collaboration} {PHENIX}),\ }\bibfield
  {title} {\bibinfo {title} {{Suppression pattern of neutral pions at high
  transverse momentum in Au$+$Au collisions at $\sqrt{s_{NN}}=$ 200 GeV and
  constraints on medium transport coefficients}},\ }\href
  {https://doi.org/10.1103/PhysRevLett.101.232301} {\bibfield  {journal}
  {\bibinfo  {journal} {Phys. Rev. Lett.}\ }\textbf {\bibinfo {volume} {101}},\
  \bibinfo {pages} {232301} (\bibinfo {year} {2008})},\ \Eprint
  {https://arxiv.org/abs/0801.4020} {arXiv:0801.4020 [nucl-ex]} \BibitemShut
  {NoStop}%
\bibitem [{\citenamefont {Adare}\ \emph {et~al.}(2013)\citenamefont {Adare}
  \emph {et~al.}}]{Adare:2012wg}%
  \BibitemOpen
  \bibfield  {author} {\bibinfo {author} {\bibfnamefont {A.}~\bibnamefont
  {Adare}} \emph {et~al.} (\bibinfo {collaboration} {PHENIX}),\ }\bibfield
  {title} {\bibinfo {title} {{Neutral pion production with respect to
  centrality and reaction plane in Au$+$Au collisions at $\sqrt{s_{NN}}$=200
  GeV}},\ }\href {https://doi.org/10.1103/PhysRevC.87.034911} {\bibfield
  {journal} {\bibinfo  {journal} {Phys. Rev. C}\ }\textbf {\bibinfo {volume}
  {87}},\ \bibinfo {pages} {034911} (\bibinfo {year} {2013})},\ \Eprint
  {https://arxiv.org/abs/1208.2254} {arXiv:1208.2254 [nucl-ex]} \BibitemShut
  {NoStop}%
\bibitem [{\citenamefont {Chatrchyan}\ \emph {et~al.}(2012)\citenamefont
  {Chatrchyan} \emph {et~al.}}]{CMS:2012aa}%
  \BibitemOpen
  \bibfield  {author} {\bibinfo {author} {\bibfnamefont {S.}~\bibnamefont
  {Chatrchyan}} \emph {et~al.} (\bibinfo {collaboration} {CMS}),\ }\bibfield
  {title} {\bibinfo {title} {{Study of high-pT charged particle suppression in
  PbPb compared to $pp$ collisions at $\sqrt{s_{NN}}=2.76$ TeV}},\ }\href
  {https://doi.org/10.1140/epjc/s10052-012-1945-x} {\bibfield  {journal}
  {\bibinfo  {journal} {Eur. Phys. J. C}\ }\textbf {\bibinfo {volume} {72}},\
  \bibinfo {pages} {1945} (\bibinfo {year} {2012})},\ \Eprint
  {https://arxiv.org/abs/1202.2554} {arXiv:1202.2554 [nucl-ex]} \BibitemShut
  {NoStop}%
\bibitem [{\citenamefont {Abelev}\ \emph {et~al.}(2013)\citenamefont {Abelev}
  \emph {et~al.}}]{Abelev:2012hxa}%
  \BibitemOpen
  \bibfield  {author} {\bibinfo {author} {\bibfnamefont {B.}~\bibnamefont
  {Abelev}} \emph {et~al.} (\bibinfo {collaboration} {ALICE}),\ }\bibfield
  {title} {\bibinfo {title} {{Centrality Dependence of Charged Particle
  Production at Large Transverse Momentum in Pb--Pb Collisions at
  $\sqrt{s_{\rm{NN}}} = 2.76$ TeV}},\ }\href
  {https://doi.org/10.1016/j.physletb.2013.01.051} {\bibfield  {journal}
  {\bibinfo  {journal} {Phys. Lett. B}\ }\textbf {\bibinfo {volume} {720}},\
  \bibinfo {pages} {52} (\bibinfo {year} {2013})},\ \Eprint
  {https://arxiv.org/abs/1208.2711} {arXiv:1208.2711 [hep-ex]} \BibitemShut
  {NoStop}%
\bibitem [{\citenamefont {Aad}\ \emph {et~al.}(2015)\citenamefont {Aad} \emph
  {et~al.}}]{Aad:2015wga}%
  \BibitemOpen
  \bibfield  {author} {\bibinfo {author} {\bibfnamefont {G.}~\bibnamefont
  {Aad}} \emph {et~al.} (\bibinfo {collaboration} {ATLAS}),\ }\bibfield
  {title} {\bibinfo {title} {{Measurement of charged-particle spectra in Pb+Pb
  collisions at $\sqrt{{s}_\mathsf{{NN}}} = 2.76$ TeV with the ATLAS detector
  at the LHC}},\ }\href {https://doi.org/10.1007/JHEP09(2015)050} {\bibfield
  {journal} {\bibinfo  {journal} {JHEP}\ }\textbf {\bibinfo {volume} {09}},\
  \bibinfo {pages} {050}},\ \Eprint {https://arxiv.org/abs/1504.04337}
  {arXiv:1504.04337 [hep-ex]} \BibitemShut {NoStop}%
\bibitem [{\citenamefont {Khachatryan}\ \emph {et~al.}(2017)\citenamefont
  {Khachatryan} \emph {et~al.}}]{Khachatryan:2016odn}%
  \BibitemOpen
  \bibfield  {author} {\bibinfo {author} {\bibfnamefont {V.}~\bibnamefont
  {Khachatryan}} \emph {et~al.} (\bibinfo {collaboration} {CMS}),\ }\bibfield
  {title} {\bibinfo {title} {{Charged-particle nuclear modification factors in
  PbPb and pPb collisions at $ \sqrt{s_{\mathrm{N}\;\mathrm{N}}}=5.02 $ TeV}},\
  }\href {https://doi.org/10.1007/JHEP04(2017)039} {\bibfield  {journal}
  {\bibinfo  {journal} {JHEP}\ }\textbf {\bibinfo {volume} {04}},\ \bibinfo
  {pages} {039}},\ \Eprint {https://arxiv.org/abs/1611.01664} {arXiv:1611.01664
  [nucl-ex]} \BibitemShut {NoStop}%
\bibitem [{\citenamefont {Acharya}\ \emph {et~al.}(2018)\citenamefont {Acharya}
  \emph {et~al.}}]{Acharya:2018qsh}%
  \BibitemOpen
  \bibfield  {author} {\bibinfo {author} {\bibfnamefont {S.}~\bibnamefont
  {Acharya}} \emph {et~al.} (\bibinfo {collaboration} {ALICE}),\ }\bibfield
  {title} {\bibinfo {title} {{Transverse momentum spectra and nuclear
  modification factors of charged particles in pp, p-Pb and Pb-Pb collisions at
  the LHC}},\ }\href {https://doi.org/10.1007/JHEP11(2018)013} {\bibfield
  {journal} {\bibinfo  {journal} {JHEP}\ }\textbf {\bibinfo {volume} {11}},\
  \bibinfo {pages} {013}},\ \Eprint {https://arxiv.org/abs/1802.09145}
  {arXiv:1802.09145 [nucl-ex]} \BibitemShut {NoStop}%
\bibitem [{\citenamefont {Abelev}\ \emph {et~al.}(2010)\citenamefont {Abelev}
  \emph {et~al.}}]{Abelev:2009gu}%
  \BibitemOpen
  \bibfield  {author} {\bibinfo {author} {\bibfnamefont {B.~I.}\ \bibnamefont
  {Abelev}} \emph {et~al.} (\bibinfo {collaboration} {STAR}),\ }\bibfield
  {title} {\bibinfo {title} {{Studying Parton Energy Loss in Heavy-Ion
  Collisions via Direct-Photon and Charged-Particle Azimuthal Correlations}},\
  }\href {https://doi.org/10.1103/PhysRevC.82.034909} {\bibfield  {journal}
  {\bibinfo  {journal} {Phys. Rev. C}\ }\textbf {\bibinfo {volume} {82}},\
  \bibinfo {pages} {034909} (\bibinfo {year} {2010})},\ \Eprint
  {https://arxiv.org/abs/0912.1871} {arXiv:0912.1871 [nucl-ex]} \BibitemShut
  {NoStop}%
\bibitem [{\citenamefont {Adamczyk}\ \emph {et~al.}(2016)\citenamefont
  {Adamczyk} \emph {et~al.}}]{STAR:2016jdz}%
  \BibitemOpen
  \bibfield  {author} {\bibinfo {author} {\bibfnamefont {L.}~\bibnamefont
  {Adamczyk}} \emph {et~al.} (\bibinfo {collaboration} {STAR}),\ }\bibfield
  {title} {\bibinfo {title} {{Jet-like Correlations with Direct-Photon and
  Neutral-Pion Triggers at $\sqrt{s_{_{NN}}} = 200$ GeV}},\ }\href
  {https://doi.org/10.1016/j.physletb.2016.07.046} {\bibfield  {journal}
  {\bibinfo  {journal} {Phys. Lett. B}\ }\textbf {\bibinfo {volume} {760}},\
  \bibinfo {pages} {689} (\bibinfo {year} {2016})},\ \Eprint
  {https://arxiv.org/abs/1604.01117} {arXiv:1604.01117 [nucl-ex]} \BibitemShut
  {NoStop}%
\bibitem [{\citenamefont {Aamodt}\ \emph {et~al.}(2012)\citenamefont {Aamodt}
  \emph {et~al.}}]{Aamodt:2011vg}%
  \BibitemOpen
  \bibfield  {author} {\bibinfo {author} {\bibfnamefont {K.}~\bibnamefont
  {Aamodt}} \emph {et~al.} (\bibinfo {collaboration} {ALICE}),\ }\bibfield
  {title} {\bibinfo {title} {{Particle-yield modification in jet-like azimuthal
  di-hadron correlations in Pb-Pb collisions at $\sqrt{s_{NN}} = 2.76$ TeV}},\
  }\href {https://doi.org/10.1103/PhysRevLett.108.092301} {\bibfield  {journal}
  {\bibinfo  {journal} {Phys. Rev. Lett.}\ }\textbf {\bibinfo {volume} {108}},\
  \bibinfo {pages} {092301} (\bibinfo {year} {2012})},\ \Eprint
  {https://arxiv.org/abs/1110.0121} {arXiv:1110.0121 [nucl-ex]} \BibitemShut
  {NoStop}%
\bibitem [{\citenamefont {Adam}\ \emph {et~al.}(2016)\citenamefont {Adam} \emph
  {et~al.}}]{Adam:2016xbp}%
  \BibitemOpen
  \bibfield  {author} {\bibinfo {author} {\bibfnamefont {J.}~\bibnamefont
  {Adam}} \emph {et~al.} (\bibinfo {collaboration} {ALICE}),\ }\bibfield
  {title} {\bibinfo {title} {{Jet-like correlations with neutral pion triggers
  in pp and central Pb\textendash{}Pb collisions at 2.76 TeV}},\ }\href
  {https://doi.org/10.1016/j.physletb.2016.10.048} {\bibfield  {journal}
  {\bibinfo  {journal} {Phys. Lett. B}\ }\textbf {\bibinfo {volume} {763}},\
  \bibinfo {pages} {238} (\bibinfo {year} {2016})},\ \Eprint
  {https://arxiv.org/abs/1608.07201} {arXiv:1608.07201 [nucl-ex]} \BibitemShut
  {NoStop}%
\bibitem [{\citenamefont {Conway}(2013)}]{Conway:2013xaa}%
  \BibitemOpen
  \bibfield  {author} {\bibinfo {author} {\bibfnamefont {R.}~\bibnamefont
  {Conway}} (\bibinfo {collaboration} {CMS}),\ }\bibfield  {title} {\bibinfo
  {title} {{Very High-$p_{T}$ triggered dihadron correlations in PbPb
  collisions at 2.76 TeV with CMS}},\ }\href
  {https://doi.org/10.1016/j.nuclphysa.2013.02.046} {\bibfield  {journal}
  {\bibinfo  {journal} {Nucl. Phys. A}\ }\textbf {\bibinfo {volume}
  {904-905}},\ \bibinfo {pages} {451c} (\bibinfo {year} {2013})}\BibitemShut
  {NoStop}%
\bibitem [{\citenamefont {Tachibana}\ \emph {et~al.}(2017)\citenamefont
  {Tachibana}, \citenamefont {Chang},\ and\ \citenamefont
  {Qin}}]{PhysRevC.95.044909}%
  \BibitemOpen
  \bibfield  {author} {\bibinfo {author} {\bibfnamefont {Y.}~\bibnamefont
  {Tachibana}}, \bibinfo {author} {\bibfnamefont {N.-B.}\ \bibnamefont
  {Chang}},\ and\ \bibinfo {author} {\bibfnamefont {G.-Y.}\ \bibnamefont
  {Qin}},\ }\bibfield  {title} {\bibinfo {title} {Full jet in quark-gluon
  plasma with hydrodynamic medium response},\ }\href
  {https://doi.org/10.1103/PhysRevC.95.044909} {\bibfield  {journal} {\bibinfo
  {journal} {Phys. Rev. C}\ }\textbf {\bibinfo {volume} {95}},\ \bibinfo
  {pages} {044909} (\bibinfo {year} {2017})}\BibitemShut {NoStop}%
\bibitem [{\citenamefont {Chen}\ \emph {et~al.}(2018)\citenamefont {Chen},
  \citenamefont {Cao}, \citenamefont {Luo}, \citenamefont {Pang},\ and\
  \citenamefont {Wang}}]{Chen:2017zte}%
  \BibitemOpen
  \bibfield  {author} {\bibinfo {author} {\bibfnamefont {W.}~\bibnamefont
  {Chen}}, \bibinfo {author} {\bibfnamefont {S.}~\bibnamefont {Cao}}, \bibinfo
  {author} {\bibfnamefont {T.}~\bibnamefont {Luo}}, \bibinfo {author}
  {\bibfnamefont {L.-G.}\ \bibnamefont {Pang}},\ and\ \bibinfo {author}
  {\bibfnamefont {X.-N.}\ \bibnamefont {Wang}},\ }\bibfield  {title} {\bibinfo
  {title} {{Effects of jet-induced medium excitation in $\gamma$-hadron
  correlation in A+A collisions}},\ }\href
  {https://doi.org/10.1016/j.physletb.2017.12.015} {\bibfield  {journal}
  {\bibinfo  {journal} {Phys. Lett. B}\ }\textbf {\bibinfo {volume} {777}},\
  \bibinfo {pages} {86} (\bibinfo {year} {2018})},\ \Eprint
  {https://arxiv.org/abs/1704.03648} {arXiv:1704.03648 [nucl-th]} \BibitemShut
  {NoStop}%
\bibitem [{\citenamefont {Kunnawalkam~Elayavalli}\ and\ \citenamefont
  {Zapp}(2017)}]{KunnawalkamElayavalli:2017hxo}%
  \BibitemOpen
  \bibfield  {author} {\bibinfo {author} {\bibfnamefont {R.}~\bibnamefont
  {Kunnawalkam~Elayavalli}}\ and\ \bibinfo {author} {\bibfnamefont {K.~C.}\
  \bibnamefont {Zapp}},\ }\bibfield  {title} {\bibinfo {title} {{Medium
  response in JEWEL and its impact on jet shape observables in heavy ion
  collisions}},\ }\href {https://doi.org/10.1007/JHEP07(2017)141} {\bibfield
  {journal} {\bibinfo  {journal} {JHEP}\ }\textbf {\bibinfo {volume} {07}},\
  \bibinfo {pages} {141}},\ \Eprint {https://arxiv.org/abs/1707.01539}
  {arXiv:1707.01539 [hep-ph]} \BibitemShut {NoStop}%
\bibitem [{\citenamefont {Milhano}\ \emph {et~al.}(2018)\citenamefont
  {Milhano}, \citenamefont {Wiedemann},\ and\ \citenamefont
  {Zapp}}]{Milhano:2017nzm}%
  \BibitemOpen
  \bibfield  {author} {\bibinfo {author} {\bibfnamefont {G.}~\bibnamefont
  {Milhano}}, \bibinfo {author} {\bibfnamefont {U.~A.}\ \bibnamefont
  {Wiedemann}},\ and\ \bibinfo {author} {\bibfnamefont {K.~C.}\ \bibnamefont
  {Zapp}},\ }\bibfield  {title} {\bibinfo {title} {{Sensitivity of jet
  substructure to jet-induced medium response}},\ }\href
  {https://doi.org/10.1016/j.physletb.2018.01.029} {\bibfield  {journal}
  {\bibinfo  {journal} {Phys. Lett. B}\ }\textbf {\bibinfo {volume} {779}},\
  \bibinfo {pages} {409} (\bibinfo {year} {2018})},\ \Eprint
  {https://arxiv.org/abs/1707.04142} {arXiv:1707.04142 [hep-ph]} \BibitemShut
  {NoStop}%
\bibitem [{\citenamefont {He}\ \emph {et~al.}(2019)\citenamefont {He},
  \citenamefont {Cao}, \citenamefont {Chen}, \citenamefont {Luo}, \citenamefont
  {Pang},\ and\ \citenamefont {Wang}}]{He:2018xjv}%
  \BibitemOpen
  \bibfield  {author} {\bibinfo {author} {\bibfnamefont {Y.}~\bibnamefont
  {He}}, \bibinfo {author} {\bibfnamefont {S.}~\bibnamefont {Cao}}, \bibinfo
  {author} {\bibfnamefont {W.}~\bibnamefont {Chen}}, \bibinfo {author}
  {\bibfnamefont {T.}~\bibnamefont {Luo}}, \bibinfo {author} {\bibfnamefont
  {L.-G.}\ \bibnamefont {Pang}},\ and\ \bibinfo {author} {\bibfnamefont
  {X.-N.}\ \bibnamefont {Wang}},\ }\bibfield  {title} {\bibinfo {title}
  {{Interplaying mechanisms behind single inclusive jet suppression in
  heavy-ion collisions}},\ }\href {https://doi.org/10.1103/PhysRevC.99.054911}
  {\bibfield  {journal} {\bibinfo  {journal} {Phys. Rev. C}\ }\textbf {\bibinfo
  {volume} {99}},\ \bibinfo {pages} {054911} (\bibinfo {year} {2019})},\
  \Eprint {https://arxiv.org/abs/1809.02525} {arXiv:1809.02525 [nucl-th]}
  \BibitemShut {NoStop}%
\bibitem [{\citenamefont {Owens}(1987)}]{Owens:1986mp}%
  \BibitemOpen
  \bibfield  {author} {\bibinfo {author} {\bibfnamefont {J.~F.}\ \bibnamefont
  {Owens}},\ }\bibfield  {title} {\bibinfo {title} {{Large Momentum Transfer
  Production of Direct Photons, Jets, and Particles}},\ }\href
  {https://doi.org/10.1103/RevModPhys.59.465} {\bibfield  {journal} {\bibinfo
  {journal} {Rev. Mod. Phys.}\ }\textbf {\bibinfo {volume} {59}},\ \bibinfo
  {pages} {465} (\bibinfo {year} {1987})}\BibitemShut {NoStop}%
\bibitem [{\citenamefont {Hou}\ \emph {et~al.}(2017)\citenamefont {Hou},
  \citenamefont {Dulat}, \citenamefont {Gao}, \citenamefont {Guzzi},
  \citenamefont {Huston}, \citenamefont {Nadolsky}, \citenamefont {Pumplin},
  \citenamefont {Schmidt}, \citenamefont {Stump},\ and\ \citenamefont
  {Yuan}}]{Hou:2016nqm}%
  \BibitemOpen
  \bibfield  {author} {\bibinfo {author} {\bibfnamefont {T.-J.}\ \bibnamefont
  {Hou}}, \bibinfo {author} {\bibfnamefont {S.}~\bibnamefont {Dulat}}, \bibinfo
  {author} {\bibfnamefont {J.}~\bibnamefont {Gao}}, \bibinfo {author}
  {\bibfnamefont {M.}~\bibnamefont {Guzzi}}, \bibinfo {author} {\bibfnamefont
  {J.}~\bibnamefont {Huston}}, \bibinfo {author} {\bibfnamefont
  {P.}~\bibnamefont {Nadolsky}}, \bibinfo {author} {\bibfnamefont
  {J.}~\bibnamefont {Pumplin}}, \bibinfo {author} {\bibfnamefont
  {C.}~\bibnamefont {Schmidt}}, \bibinfo {author} {\bibfnamefont
  {D.}~\bibnamefont {Stump}},\ and\ \bibinfo {author} {\bibfnamefont {C.~P.}\
  \bibnamefont {Yuan}},\ }\bibfield  {title} {\bibinfo {title} {{CTEQ-TEA
  parton distribution functions and HERA Run I and II combined data}},\ }\href
  {https://doi.org/10.1103/PhysRevD.95.034003} {\bibfield  {journal} {\bibinfo
  {journal} {Phys. Rev. D}\ }\textbf {\bibinfo {volume} {95}},\ \bibinfo
  {pages} {034003} (\bibinfo {year} {2017})},\ \Eprint
  {https://arxiv.org/abs/1609.07968} {arXiv:1609.07968 [hep-ph]} \BibitemShut
  {NoStop}%
\bibitem [{\citenamefont {Kniehl}\ \emph {et~al.}(2000)\citenamefont {Kniehl},
  \citenamefont {Kramer},\ and\ \citenamefont {Potter}}]{Kniehl:2000fe}%
  \BibitemOpen
  \bibfield  {author} {\bibinfo {author} {\bibfnamefont {B.~A.}\ \bibnamefont
  {Kniehl}}, \bibinfo {author} {\bibfnamefont {G.}~\bibnamefont {Kramer}},\
  and\ \bibinfo {author} {\bibfnamefont {B.}~\bibnamefont {Potter}},\
  }\bibfield  {title} {\bibinfo {title} {{Fragmentation functions for pions,
  kaons, and protons at next-to-leading order}},\ }\href
  {https://doi.org/10.1016/S0550-3213(00)00303-5} {\bibfield  {journal}
  {\bibinfo  {journal} {Nucl. Phys. B}\ }\textbf {\bibinfo {volume} {582}},\
  \bibinfo {pages} {514} (\bibinfo {year} {2000})},\ \Eprint
  {https://arxiv.org/abs/hep-ph/0010289} {arXiv:hep-ph/0010289} \BibitemShut
  {NoStop}%
\bibitem [{\citenamefont {Eskola}\ \emph {et~al.}(2017)\citenamefont {Eskola},
  \citenamefont {Paakkinen}, \citenamefont {Paukkunen},\ and\ \citenamefont
  {Salgado}}]{Eskola:2016oht}%
  \BibitemOpen
  \bibfield  {author} {\bibinfo {author} {\bibfnamefont {K.~J.}\ \bibnamefont
  {Eskola}}, \bibinfo {author} {\bibfnamefont {P.}~\bibnamefont {Paakkinen}},
  \bibinfo {author} {\bibfnamefont {H.}~\bibnamefont {Paukkunen}},\ and\
  \bibinfo {author} {\bibfnamefont {C.~A.}\ \bibnamefont {Salgado}},\
  }\bibfield  {title} {\bibinfo {title} {{EPPS16: Nuclear parton distributions
  with LHC data}},\ }\href {https://doi.org/10.1140/epjc/s10052-017-4725-9}
  {\bibfield  {journal} {\bibinfo  {journal} {Eur. Phys. J. C}\ }\textbf
  {\bibinfo {volume} {77}},\ \bibinfo {pages} {163} (\bibinfo {year} {2017})},\
  \Eprint {https://arxiv.org/abs/1612.05741} {arXiv:1612.05741 [hep-ph]}
  \BibitemShut {NoStop}%
\bibitem [{\citenamefont {Wang}(2000)}]{Wang:1998ww}%
  \BibitemOpen
  \bibfield  {author} {\bibinfo {author} {\bibfnamefont {X.-N.}\ \bibnamefont
  {Wang}},\ }\bibfield  {title} {\bibinfo {title} {{Systematic study of high
  $p_{T}$ hadron spectra in $p p$, $p$ A and A A collisions from SPS to RHIC
  energies}},\ }\href {https://doi.org/10.1103/PhysRevC.61.064910} {\bibfield
  {journal} {\bibinfo  {journal} {Phys. Rev. C}\ }\textbf {\bibinfo {volume}
  {61}},\ \bibinfo {pages} {064910} (\bibinfo {year} {2000})},\ \Eprint
  {https://arxiv.org/abs/nucl-th/9812021} {arXiv:nucl-th/9812021} \BibitemShut
  {NoStop}%
\bibitem [{\citenamefont {Hirano}\ and\ \citenamefont
  {Nara}(2004)}]{Hirano:2003pw}%
  \BibitemOpen
  \bibfield  {author} {\bibinfo {author} {\bibfnamefont {T.}~\bibnamefont
  {Hirano}}\ and\ \bibinfo {author} {\bibfnamefont {Y.}~\bibnamefont {Nara}},\
  }\bibfield  {title} {\bibinfo {title} {{Interplay between soft and hard
  hadronic components for identified hadrons in relativistic heavy ion
  collisions at RHIC}},\ }\href {https://doi.org/10.1103/PhysRevC.69.034908}
  {\bibfield  {journal} {\bibinfo  {journal} {Phys. Rev. C}\ }\textbf {\bibinfo
  {volume} {69}},\ \bibinfo {pages} {034908} (\bibinfo {year} {2004})},\
  \Eprint {https://arxiv.org/abs/nucl-th/0307015} {arXiv:nucl-th/0307015}
  \BibitemShut {NoStop}%
\bibitem [{\citenamefont {Xie}\ \emph {et~al.}()\citenamefont {Xie},
  \citenamefont {Ke}, \citenamefont {Han-Zhong},\ and\ \citenamefont
  {Wang}}]{longpaper}%
  \BibitemOpen
  \bibfield  {author} {\bibinfo {author} {\bibfnamefont {M.}~\bibnamefont
  {Xie}}, \bibinfo {author} {\bibfnamefont {W.}~\bibnamefont {Ke}}, \bibinfo
  {author} {\bibfnamefont {Z.}~\bibnamefont {Han-Zhong}},\ and\ \bibinfo
  {author} {\bibfnamefont {X.-N.}\ \bibnamefont {Wang}},\ }\href@noop {}
  {\bibinfo {title} {{In preparation}}}\BibitemShut {NoStop}%
\bibitem [{\citenamefont {Wang}(2004)}]{Wang:2004yv}%
  \BibitemOpen
  \bibfield  {author} {\bibinfo {author} {\bibfnamefont {X.-N.}\ \bibnamefont
  {Wang}},\ }\bibfield  {title} {\bibinfo {title} {{Energy dependence of jet
  quenching and life-time of the dense matter in high-energy heavy-ion
  collisions}},\ }\href {https://doi.org/10.1103/PhysRevC.70.031901} {\bibfield
   {journal} {\bibinfo  {journal} {Phys. Rev. C}\ }\textbf {\bibinfo {volume}
  {70}},\ \bibinfo {pages} {031901} (\bibinfo {year} {2004})},\ \Eprint
  {https://arxiv.org/abs/nucl-th/0405029} {arXiv:nucl-th/0405029} \BibitemShut
  {NoStop}%
\bibitem [{\citenamefont {Zhang}\ \emph {et~al.}(2007)\citenamefont {Zhang},
  \citenamefont {Owens}, \citenamefont {Wang},\ and\ \citenamefont
  {Wang}}]{Zhang:2007ja}%
  \BibitemOpen
  \bibfield  {author} {\bibinfo {author} {\bibfnamefont {H.}~\bibnamefont
  {Zhang}}, \bibinfo {author} {\bibfnamefont {J.~F.}\ \bibnamefont {Owens}},
  \bibinfo {author} {\bibfnamefont {E.}~\bibnamefont {Wang}},\ and\ \bibinfo
  {author} {\bibfnamefont {X.-N.}\ \bibnamefont {Wang}},\ }\bibfield  {title}
  {\bibinfo {title} {{Dihadron tomography of high-energy nuclear collisions in
  NLO pQCD}},\ }\href {https://doi.org/10.1103/PhysRevLett.98.212301}
  {\bibfield  {journal} {\bibinfo  {journal} {Phys. Rev. Lett.}\ }\textbf
  {\bibinfo {volume} {98}},\ \bibinfo {pages} {212301} (\bibinfo {year}
  {2007})},\ \Eprint {https://arxiv.org/abs/nucl-th/0701045}
  {arXiv:nucl-th/0701045} \BibitemShut {NoStop}%
\bibitem [{\citenamefont {Zhang}\ \emph {et~al.}(2009)\citenamefont {Zhang},
  \citenamefont {Owens}, \citenamefont {Wang},\ and\ \citenamefont
  {Wang}}]{Zhang:2009rn}%
  \BibitemOpen
  \bibfield  {author} {\bibinfo {author} {\bibfnamefont {H.}~\bibnamefont
  {Zhang}}, \bibinfo {author} {\bibfnamefont {J.~F.}\ \bibnamefont {Owens}},
  \bibinfo {author} {\bibfnamefont {E.}~\bibnamefont {Wang}},\ and\ \bibinfo
  {author} {\bibfnamefont {X.-N.}\ \bibnamefont {Wang}},\ }\bibfield  {title}
  {\bibinfo {title} {{Tomography of high-energy nuclear collisions with
  photon-hadron correlations}},\ }\href
  {https://doi.org/10.1103/PhysRevLett.103.032302} {\bibfield  {journal}
  {\bibinfo  {journal} {Phys. Rev. Lett.}\ }\textbf {\bibinfo {volume} {103}},\
  \bibinfo {pages} {032302} (\bibinfo {year} {2009})},\ \Eprint
  {https://arxiv.org/abs/0902.4000} {arXiv:0902.4000 [nucl-th]} \BibitemShut
  {NoStop}%
\bibitem [{\citenamefont {Pang}\ \emph {et~al.}(2012)\citenamefont {Pang},
  \citenamefont {Wang},\ and\ \citenamefont {Wang}}]{Pang:2012he}%
  \BibitemOpen
  \bibfield  {author} {\bibinfo {author} {\bibfnamefont {L.}~\bibnamefont
  {Pang}}, \bibinfo {author} {\bibfnamefont {Q.}~\bibnamefont {Wang}},\ and\
  \bibinfo {author} {\bibfnamefont {X.-N.}\ \bibnamefont {Wang}},\ }\bibfield
  {title} {\bibinfo {title} {{Effects of initial flow velocity fluctuation in
  event-by-event (3+1)D hydrodynamics}},\ }\href
  {https://doi.org/10.1103/PhysRevC.86.024911} {\bibfield  {journal} {\bibinfo
  {journal} {Phys. Rev. C}\ }\textbf {\bibinfo {volume} {86}},\ \bibinfo
  {pages} {024911} (\bibinfo {year} {2012})},\ \Eprint
  {https://arxiv.org/abs/1205.5019} {arXiv:1205.5019 [nucl-th]} \BibitemShut
  {NoStop}%
\bibitem [{\citenamefont {Pang}\ \emph {et~al.}(2015)\citenamefont {Pang},
  \citenamefont {Hatta}, \citenamefont {Wang},\ and\ \citenamefont
  {Xiao}}]{Pang:2014ipa}%
  \BibitemOpen
  \bibfield  {author} {\bibinfo {author} {\bibfnamefont {L.-G.}\ \bibnamefont
  {Pang}}, \bibinfo {author} {\bibfnamefont {Y.}~\bibnamefont {Hatta}},
  \bibinfo {author} {\bibfnamefont {X.-N.}\ \bibnamefont {Wang}},\ and\
  \bibinfo {author} {\bibfnamefont {B.-W.}\ \bibnamefont {Xiao}},\ }\bibfield
  {title} {\bibinfo {title} {{Analytical and numerical Gubser solutions of the
  second-order hydrodynamics}},\ }\href
  {https://doi.org/10.1103/PhysRevD.91.074027} {\bibfield  {journal} {\bibinfo
  {journal} {Phys. Rev. D}\ }\textbf {\bibinfo {volume} {91}},\ \bibinfo
  {pages} {074027} (\bibinfo {year} {2015})},\ \Eprint
  {https://arxiv.org/abs/1411.7767} {arXiv:1411.7767 [hep-ph]} \BibitemShut
  {NoStop}%
\bibitem [{\citenamefont {Pang}\ \emph {et~al.}(2018)\citenamefont {Pang},
  \citenamefont {Petersen},\ and\ \citenamefont {Wang}}]{Pang:2018zzo}%
  \BibitemOpen
  \bibfield  {author} {\bibinfo {author} {\bibfnamefont {L.-G.}\ \bibnamefont
  {Pang}}, \bibinfo {author} {\bibfnamefont {H.}~\bibnamefont {Petersen}},\
  and\ \bibinfo {author} {\bibfnamefont {X.-N.}\ \bibnamefont {Wang}},\
  }\bibfield  {title} {\bibinfo {title} {{Pseudorapidity distribution and
  decorrelation of anisotropic flow within the open-computing-language
  implementation CLVisc hydrodynamics}},\ }\href
  {https://doi.org/10.1103/PhysRevC.97.064918} {\bibfield  {journal} {\bibinfo
  {journal} {Phys. Rev. C}\ }\textbf {\bibinfo {volume} {97}},\ \bibinfo
  {pages} {064918} (\bibinfo {year} {2018})},\ \Eprint
  {https://arxiv.org/abs/1802.04449} {arXiv:1802.04449 [nucl-th]} \BibitemShut
  {NoStop}%
\bibitem [{\citenamefont {Moreland}\ \emph {et~al.}(2015)\citenamefont
  {Moreland}, \citenamefont {Bernhard},\ and\ \citenamefont
  {Bass}}]{Moreland:2014oya}%
  \BibitemOpen
  \bibfield  {author} {\bibinfo {author} {\bibfnamefont {J.~S.}\ \bibnamefont
  {Moreland}}, \bibinfo {author} {\bibfnamefont {J.~E.}\ \bibnamefont
  {Bernhard}},\ and\ \bibinfo {author} {\bibfnamefont {S.~A.}\ \bibnamefont
  {Bass}},\ }\bibfield  {title} {\bibinfo {title} {{Alternative ansatz to
  wounded nucleon and binary collision scaling in high-energy nuclear
  collisions}},\ }\href {https://doi.org/10.1103/PhysRevC.92.011901} {\bibfield
   {journal} {\bibinfo  {journal} {Phys. Rev. C}\ }\textbf {\bibinfo {volume}
  {92}},\ \bibinfo {pages} {011901} (\bibinfo {year} {2015})},\ \Eprint
  {https://arxiv.org/abs/1412.4708} {arXiv:1412.4708 [nucl-th]} \BibitemShut
  {NoStop}%
\bibitem [{Note1()}]{Note1}%
  \BibitemOpen
  \bibinfo {note} {An overall envelope function in the spatial rapidity is used
  to generalized the TRENTo initial condition at middle rapidity to a 3D
  distribution.}\BibitemShut {Stop}%
\bibitem [{\citenamefont {Ke}(2019)}]{Ke:2019jbh}%
  \BibitemOpen
  \bibfield  {author} {\bibinfo {author} {\bibfnamefont {W.}~\bibnamefont
  {Ke}},\ }\emph {\bibinfo {title} {{Partonic transport model application to
  heavy flavor}}},\ \href@noop {} {\bibinfo {type} {Phd dissertation}},\
  \bibinfo  {school} {Duke University} (\bibinfo {year} {2019}),\ \Eprint
  {https://arxiv.org/abs/2001.02766} {arXiv:2001.02766 [nucl-th]} \BibitemShut
  {NoStop}%
\bibitem [{Note2()}]{Note2}%
  \BibitemOpen
  \bibinfo {note} {Additional constraints can be imposed by other
  pre-processing procedures.}\BibitemShut {Stop}%
\bibitem [{\citenamefont {Bernhard}(2018)}]{Bernhard:2018hnz}%
  \BibitemOpen
  \bibfield  {author} {\bibinfo {author} {\bibfnamefont {J.~E.}\ \bibnamefont
  {Bernhard}},\ }\emph {\bibinfo {title} {{Bayesian parameter estimation for
  relativistic heavy-ion collisions}}},\ \href@noop {} {\bibinfo {type} {Phd
  dissertation}},\ \bibinfo  {school} {Duke University} (\bibinfo {year}
  {2018}),\ \Eprint {https://arxiv.org/abs/1804.06469} {arXiv:1804.06469
  [nucl-th]} \BibitemShut {NoStop}%
\end{thebibliography}%

\end{document}